# A comparative study of some analytic approximate methods for the anharmonic oscillator described by cubic – quartic potential

**Sonia, Sunita Srivastava* and Vishwamittar**

*Department of Physics, Panjab University, Chandigarh - 160014, India*

**ABSTRACT**

The improved formulations of the Lindstedt - Poincare perturbation method, the harmonic balance method and its modifications using the rational functions, the energy balance method and the two-terms Fourier series expansion have been employed to find approximate solution to the equation of motion for the cubic – quartic potential AHO and its special cases (the cubic, the quartic, and the purely cubic – quartic potential AHOs), to assess the relative merits and demerits of these techniques. Two new rational functions as approximate solution for the harmonic balance method have been proposed and the rational functions have been used as trial functions in the energy balance method for the first time. The time periods $T_+$ (for $q(t) > 0$), $T_-$ (for $q(t) < 0$) and the total period T so determined have been compared with the corresponding exact values for an extensive range of the AHO parameter values. The results have been thoroughly analysed in terms of their percentage difference from the relevant exact time period. It is concluded that the methods based on the Lindstedt - Poincare perturbation approach do not lead to reliable results for $T_+$ and $T_-$, which makes their applicability for the asymmetric potential AHOs doubtful. The harmonic balance technique together with its different variants is the most successful, while the Fourier series approach also yields commendably accurate results. The outcome of the third-order energy balance method is reasonably good though the usage of the rational functions for this has not been encouraging.

## 1. Introduction

In view of the smorgasbord of applications of the anharmonic oscillators (AHOs) with strong non-linearity, various researchers have formulated a number of innovative methods to determine approximate solution for their equation of motion in one-dimension and thereby to find an expression for the frequency or time period in terms of amplitude of oscillations and other parameters characterizing the oscillator. Most of them have elaborated their techniques by considering the Duffing AHO with symmetric potential

$$V(q) = k_2 q^2 + k_4 q^4 \quad (1)$$

as typical example and comparing the output with the results of one or two other earlier developed methods and the exact numerical computations.

Prompted by the availability of significantly large variety of approaches, we undertook the task of examining the relative merits and demerits of some of the techniques involving trigonometric functions in the approximate solutions and to look into their inner relationships. For this purpose, we have used the cubic - quartic potential AHO,

$$V(q) = m\left[\frac{1}{2}\omega^2 q^2 + \frac{1}{3}aq^3 + \frac{1}{4}bq^4\right], \tag{2}$$

as the main test case for a wide range of the parameter values. The choice of this asymmetric AHO has been mainly guided by the fact that the solution of its equation of motion

$$\ddot{q}\,(t) + \omega^2 q(t) + aq^2(t) + bq^3(t) = 0 \tag{3}$$

is rendered more arduous by the difference in the behaviour for the positive and negative values of q(t). The asymmetry of the potential has indeed proved to be very crucial in this investigation. Besides, this potential is important in its own right as it has been successfully used for analysing the vibration spectra of molecules, understanding the thermal expansion and heat capacity of one dimensional crystals, elucidation of thermal transport in one-dimensional chain, describing the nonlinear free vibrations of thin laminated plates and interpreting some features of the plasma oscillations which play a crucial role in plasma processing [1-4]. Furthermore, this potential reduces to a cubic potential AHO for b = 0 and to a quartic AHO for a = 0. The former leads to the so called eardrum equation while the latter is the Duffing AHO. The purely cubic - quartic potential AHO can be also treated as a special case of the potential given by (2) with ω = 0. It may be also pointed out that the dimensionless equation along the axial direction for a nonlinear ion trap wherein the quadrupole, the hexapole and the octopole terms are taken into account, looks like Eq. (3) [5].

Eq. (3) has been solved by Hu [6] employing the first-order harmonic balance method (HBM) for ω = b = 1 and the numerical calculations have been reported for a = 1 and different values of the amplitude of oscillation. He has also used the same technique for the special case of b = 0 (the eardrum equation) and ω = 1 giving numerical results for a = 1 and some values of the amplitude [7]. In addition, Elias-Zuniga [8] put forward a Jacobian elliptic function based ansatz to find the exact solution for a similar equation and derived conditions for its existence.

The choice of the techniques selected for the present study has been guided by their reported accuracy for the quartic potential AHO, wherein at most three terms have been retained in the expansion of the approximate solution. The methods investigated can be grouped as modified Lindstedt - Poincare perturbation methods (LPM) [9-12], the HBM along with the rational function based approach [13-18], the energy balance method (EBM) [19,20], and the two-terms Fourier series expansion (FSE) [21,22]. We have also put forward two new higher-order rational functions for the HBM. Moreover, the rational functions have been used for the first time in the EBM. In spite of the fact that homotopy analysis approach [23,24] is a very general analytical tool for solving nonlinear differential equations, we have refrained from its inclusion in this work because of the detailed analysis presented by Chen et al [25] and Liao's comment about its requirement of high performance computers [24].

## 2. General aspects of the cubic - quartic potential

*2.1. Separatrices*

For positive values of the AHO parameters $\omega^2$, a and b, the cubic - quartic potential, Eq.(2), has $q_0 = 0$ as a stable minimum while

$$q_{\pm} = \frac{-a \pm (a^2 - 4\omega^2 b)^{\frac{1}{2}}}{2b} \quad (4)$$

give two extrema. If $a^2 < 4\omega^2 b$, then $q_{\pm}$ become complex implying that the extrema are non-existent and the potential is a single-well like a quartic AHO for which a = 0. On the other hand, $q_{\pm}$ is realizable only when $a^2 \geq 4\omega^2 b$. The equality corresponds to the coinciding values of $q_+$ and $q_-$, namely, $q_{\pm} = -a/2b$. For this, $d^2V/dq^2$ is zero so that the point is an inflection. However, dV/dq is negative for both q infinitesimally smaller or greater than $q_{\pm}$ meaning thereby that –a/2b is not a local extremum.

The case $a^2 > 4\omega^2 b$ pertains to distinct values of $q_+$ and $q_-$ and the former, which itself is negative, is a local maximum. Therefore, if the energy of the particle is such that it exceeds the potential energy $V(q = q_+)$, the orbit will be no longer bound. Consequently, for an arbitrary initial amplitude C, the orbits of motion are homoclinic when V(q = 0) = 0 < V(q = C) < V(q = $q_+$). Substituting q = C and q = $q_+$ from Eq. (4) into Eq. (2), together with V(q = C) < V(q = $q_+$), after some simple algebra leads to the condition

$$12\omega^2 b^3 c^2 + 8ab^3 c^3 + 6b^4 c^4 < -a^4 + 6\omega^2 b(a^2 - \omega^2 b) + a(a^2 - 4\omega^2 b)^{3/2}. \quad (5)$$

Otherwise, the orbits will be heteroclinic. Thus, the two terms in (5) with sign of equality define the separatrices for the cubic - quartic AHO.

The preceding consideration shows that if $\omega^2$, a and b are so chosen that $a^2 < 4\omega^2 b$, then any arbitrary value of C can be used for calculations. However, when $a^2 > 4\omega^2 b$, C has to be taken in accordance with the inequality in (5). It may be mentioned that Hu [6] studied the AHO having ω = a = b = 1 so that the condition $a^2 < 4\omega^2 b$ is satisfied and C could be picked up without a constraint.

Next, in the case of a cubic potential AHO (b = 0), $q_0 = 0$ is a stable minimum while $q_0 = -\omega^2/a$ is a maximum. Therefore, for the motion to be bound or the orbit to be homoclinic, the turning point for q < 0 must have value more than $-\omega^2/a$. Here, the condition for the orbits to be homoclinic turns out to be $aC < \omega^2/2$. The orbits will be heteroclinic for $aC > \omega^2/2$ while $aC = \omega^2/2$ gives the separatrices. This is the same result as in [21] where ω = 1.

*2.2. Relationship between amplitudes for q(t) > 0 and q(t) < 0*

The presence of cubic term in the expression for V(q) renders its plot less steep for q < 0 and more for q > 0 for a specific value of energy (E). This, in turn, makes the amplitude of the oscillation to have larger magnitude for q < 0 as compared to that for q > 0 as also pointed out by Hu [6]. Thus, taking the oscillations to be confined to the asymmetric limits [-D, C], where D > C > 0, we have for the turning points,

$$\frac{1}{2}\omega^2 D^2 - \frac{1}{3}aD^3 + \frac{1}{4}bD^4 = \frac{1}{2}\omega^2 C^2 + \frac{1}{3}aC^3 + \frac{1}{4}bC^4. \quad (6)$$

This finally leads to

$$D = \frac{C}{3} + \frac{4a}{9b} - \frac{4\eta}{3b}\left(1 + \frac{\zeta}{\eta^2}\right), \quad (7)$$

where

$$\eta = \left[\frac{r + (r^2 - 4\zeta^3)^{\frac{1}{2}}}{2}\right]^{\frac{1}{3}}, \qquad \zeta = \left(\frac{a}{3} + \frac{bC}{4}\right)^2 - \frac{3b}{4}\left(\frac{\omega^2}{2} + \frac{aC}{3} + \frac{bC^2}{4}\right),$$

and

$$r = -2\left(\frac{a}{3} + \frac{bC}{4}\right)^3 + \frac{9b}{4}\left(\frac{\omega^2}{2} + \frac{aC}{3} + \frac{bC^2}{4}\right)\left(\frac{a}{3} + \frac{bC}{4}\right) - \frac{27b^2C}{16}\left(\frac{\omega^2}{2} + \frac{aC}{3} + \frac{bC^2}{4}\right). \quad (8)$$

Therefore, for a particular value of C, whether chosen arbitrarily or as per the requirement of condition in (5), the corresponding D can be determined from Eq. (7). Once again, our expression for $\omega = 1$ corresponds to the earlier result found by Hu [6].

For a cubic AHO, Eq.(6) reduces to

$$\frac{1}{2}\omega^2 D^2 - \frac{1}{3}aD^3 = \frac{1}{2}\omega^2 C^2 + \frac{1}{3}aC^3. \quad (9)$$

The acceptable value of D, which is such that $D \to 0$ when $C \to 0$ for $E \to 0$, is given by

$$D = \frac{1}{4a}\left[(3\omega^2 + 2aC) - 3\left\{\omega^4 - \frac{4}{3}aC(\omega^2 + aC)\right\}^{\frac{1}{2}}\right], \quad (10)$$

and its special case for $\omega = 1$ has been already reported by Hu[7]. From this expression, we note that for D to be real $3\omega^4 > 4aC(\omega^2 + aC)$ which on solving for aC gives $aC < \omega^2/2$, as obtained for the orbit to be homoclinic.

*2.3. Exact time-period*

For a conservative system, the total energy $E = \frac{m}{2}\left(\frac{dq}{dt}\right)^2 + V(q)$ is a constant of motion and for the trajectory confined to [-D, C] in the cubic - quartic potential, this leads to the exact time-period of oscillations as

$$T^{ex} = T_+^{ex} + T_-^{ex}, \quad (11)$$

where

$$T_+^{ex} = \int_0^C \frac{\sqrt{2}dq}{\left[\frac{\omega^2}{2}(C^2 - q^2) + \frac{a}{3}(C^3 - q^3) + \frac{b}{4}(C^4 - q^4)\right]^{\frac{1}{2}}} \quad (12)$$

and

$$T_{-}^{ex} = \int_0^D \frac{\sqrt{2}dq}{\left[\frac{\omega^2}{2}(D^2-q^2)-\frac{a}{3}(D^3-q^3)+\frac{b}{4}(D^4-q^4)\right]^{\frac{1}{2}}}. \quad (13)$$

Here, $T_{+}^{ex}$ and $T_{-}^{ex}$ are the times taken to cover the trajectory with q(t) > 0 and q(t) < 0 in one oscillation, respectively. It may be noted that b = 0 in the above yields the expressions for the cubic AHO. Also for a quartic AHO which is described by the symmetric potential,

$$T^{ex} = \int_0^C \frac{2\sqrt{2}dq}{\left[\frac{\omega^2}{2}(C^2-q^2)+\frac{b}{4}(C^4-q^4)\right]^{\frac{1}{2}}}. \quad (14)$$

## 3. Lindstedt - Poincare perturbation method based techniques

In view of the fact that the LPM is one of the most commonly used techniques for finding approximate solutions to nonlinear oscillators falling in the perturbative domain, a number of its extensions have been propounded by various workers to handle the problems with strong anharmonicity. One approach has been to introduce a new parameter that remains small irrespective of magnitude of the original anharmonicity parameters [26]. This is the strategy underlying the techniques developed by Amore and Aranda [9, 10], Wu et al [11] and also by Sun et al[12], which have been used in this work.

In order to implement the LPM, we rescale time t to τ as τ = Ωt and, thus, Eq. (3) becomes

$$\Omega^2 q''(\tau) + \omega^2 q(\tau) + aq^2(\tau) + bq^3(\tau) = 0. \quad (15)$$

Here, Ω is angular frequency of the AHO, which like q(τ) too is unknown quantity. Following the procedure recommended by the above mentioned authors, we rewrite Eq. (15) in the form

$$\Omega^2 q''(\tau) + (\omega^2 + \alpha)q(\tau) = \varepsilon[\alpha q(\tau) - aq^2(\tau) - bq^3(\tau)], \quad (16)$$

which reduces to (15) for ε= 1. In this expression, α is additional linear spring or stiffness constant per unit mass and is treated as an adjustable parameter. ε is dummy perturbation parameter employed for book keeping. The main difference from the conventional LPM is that a as well as b need not be small and ultimately ε will be taken 1. It is quite crucial to choose α in the best possible way.

Guided by the experience of Amore and Aranda [9], we substitute the series expansions

$$q = q_0 + \varepsilon q_1 + \varepsilon^2 q_2 + \varepsilon^3 q_3, \quad (17)$$

$$\Omega^2 = \omega_0 + \varepsilon\omega_1 + \varepsilon^2\omega_2 + \varepsilon^3\omega_3 \quad (18)$$

into Eq. (16) and get the hierarchical linear differential equations as

$$\omega_0 q_0'' + (\omega^2 + \alpha)q_0 = 0, \quad (19)$$

$$\omega_0 q_1'' + (\omega^2 + \alpha)q_1 = -\omega_1 q_0'' + \alpha q_0 - aq_0^2 - bq_{0,}^3 \quad (20)$$

$$\omega_0 q_2'' + (\omega^2 + \alpha) q_2 = -\omega_2 q_0'' - \omega_1 q_1'' + \alpha q_1 - 2a q_0 q_1 - 3b q_0^2 q_1, \quad (21)$$

$$\omega_0 q_3'' + (\omega^2 + \alpha) q_3 = -\omega_3 q_0'' - \omega_2 q_1'' - \omega_1 q_2'' + \alpha q_2 - 2a q_0 q_2$$

$$-3b q_0 q_1^2 - 3b q_0^2 q_2 - a q_1^2. \quad (22)$$

Since the derivatives in these expressions are with respect to $\tau = \Omega t$ and $\Omega$ has been identified as the frequency of oscillations of the AHO, we demand that $\omega^2 + \alpha = \omega_0$ in the homogeneous differential equation, Eq. (19). Furthermore, in order to incorporate the fact that amplitudes of oscillation for $q(\tau) > 0$ and $q(\tau) < 0$ are, respectively, C and D, we have used the following conditions:

$$q_0(\tau = 0) = C, \qquad q_0(\tau = \pi) = -D$$

$$q_j(\tau = 0) = q_j(\tau = \pi) = 0 \qquad \text{for } j = 1, 2, 3 \quad (23)$$

$$q_k'(\tau = 0) = q_k'(\tau = \pi) = 0 \qquad \text{for } k = 0, 1, 2, 3.$$

Accordingly, the solution for Eq. (19) can be written as

$$q_0(t) = \begin{cases} C\cos(\Omega_+ t) & \text{for } q_0(t) > 0 \\ D\cos(\Omega_- t) & \text{for } q_0(t) < 0 \end{cases}. \quad (24)$$

The subscripts + and – correspond to the motion for $q(t) > 0$ and $q(t) < 0$, respectively. Substituting this into the first nonhomogeneous linear differential equation, Eq. (20), using the conditions spelled out in (23) for $j = k = 1$, and eliminating the undesired secular term in the expression for $q_1(t)$, we get

$$\omega_{1+} = \frac{3bC^2}{4} - \alpha_+, \qquad \omega_{1-} = \frac{3bD^2}{4} - \alpha_-. \quad (25)$$

Continuing the procedure for the next two differential equations, we obtain

$$\omega_{2+} = \frac{1}{384\omega_{0+}}[-320a^2C^2 + 192abC^3 - 9b^2C^4], \quad (26a)$$

$$\omega_{2-} = -\frac{1}{384\omega_{0-}}[320a^2D^2 + 192abD^3 + 9b^2D^4], \quad (26b)$$

and

$$\omega_{3+} = \frac{1}{4608\omega_{0+}^2}\begin{bmatrix} -2560a^3C^3 + 8880a^2bC^4 - 2448ab^2C^5 \\ +81b^3C^6 + 12\omega_{2+}\alpha_+ \end{bmatrix}, \quad (27a)$$

$$\omega_{3-} = \frac{1}{4608\omega_{0-}^2}\begin{bmatrix} 2560a^3D^3 + 8880a^2bD^4 + 2448ab^2D^5 \\ +81b^3D^6 + 12\omega_{2-}\alpha_- \end{bmatrix}. \quad (27b)$$

Here,

$$\omega_{0\pm} = \omega^2 + \alpha_\pm . \quad (28)$$

Thus, up to third-order in $\varepsilon$, the angular frequencies for the trajectories with $q(t) > 0$ and $q(t) < 0$ for $\varepsilon = 1$ are given by

$$\Omega_\pm^2 = \omega^2 + \alpha_\pm + \omega_{1\pm} + \omega_{2\pm} + \omega_{3\pm}. \quad (29)$$

Obviously, $\alpha_\pm$ have to be determined to make the results as accurate and reliable as possible.

*3.1. Amore et al approach*

Amore and Aranda [9,10] used the principle of minimum sensitivity to evaluate α in their treatment for the Duffing and other AHOs. Applying this procedure in the present case, we employ the condition $d\Omega_{\pm}^2/d\alpha_{\pm} = 0$ in (29) along with $\omega_{1\pm}$, $\omega_{2\pm}$ and $\omega_{3\pm}$ given by Eqs. (25) - (27). This gives us

$$\alpha_{+} = \frac{2560a^3C-8880a^2bC^2+2448ab^2C^3-81b^3C^4}{-3840a^2+2304abC-108b^2C^2}, \tag{30a}$$

and

$$\alpha_{-} = \frac{2560a^3D+8880a^2bD^2+2448ab^2D^3+81b^3D^4}{3840a^2+2304abD+108b^2D^2}. \tag{30b}$$

It may be pointed out that for a cubic potential AHO (b = 0),

$$\alpha_{+} = -\frac{2}{3}aC, \qquad \alpha_{-} = \frac{2}{3}aD. \tag{31}$$

These combined with Eqs. (25) - (27) and Eq. (29) yield

$$\Omega_{+}^2 = \omega^2 - \frac{5a^2C^2}{6\omega^2-4aC} \quad \text{and} \quad \Omega_{-}^2 = \omega^2 - \frac{5a^2D^2}{6\omega^2+4aD}. \tag{32}$$

It may be noted that both $\Omega_{+}^2$ and $\Omega_{-}^2$ are less than $\omega^2$. For the cubic AHO to be oscillatory, both of these must be positive. The former requires that

$$\omega > \left[\frac{2aC}{3}\left(1+\frac{5aC}{4\omega^2}\right)\right]^{\frac{1}{2}}, \tag{33a}$$

while the latter demands that

$$\omega > \left[-\frac{2aD}{3}\left(1-\frac{5aD}{4\omega^2}\right)\right]^{\frac{1}{2}}. \tag{33b}$$

These conditions can be also, respectively, expressed as $\omega > 1.142\sqrt{aC}$ and $\omega > 0.799\sqrt{aD}$. This restricts the domain of applicability of the method to the cubic potential AHO. In fact, similar conditions are also expected for the cubic - quartic AHO as the expressions for $\Omega_{\pm}^2$ involve negative terms as in Eq. (32), which are absent for the case of a quartic AHO (a = 0, D ≡ C and $\alpha_{-} \equiv \alpha_{+} = 3bC^2/4$), namely,

$$\Omega^2 = \omega^2 + \frac{3bC^2(32\omega^2+23bC^2)}{128\omega^2+96bC^2}. \tag{34}$$

*3.2. Wu and coworkers' prescription*

The recipe for determining the optimum value of the linear spring constant α, as proposed by Wu et al [11] essentially amounts to taking

$$\sum_{l=1}^{n} \omega_l = 0 \tag{35}$$

where n is the order of approximation employed for the expansion of q and $\Omega^2$ in Eqs. (17) and (18), respectively. In the case of a Duffing AHO, this approach on being implemented with LP perturbation up to second-order gave results much better than those obtained by using the prescription of Amore et al, which could be applied only if the perturbative treatment was up to third-order. Adopting this methodology for the present situation, we have up to second-order $\omega_{1\pm} + \omega_{2\pm} = 0$ . Substituting the relevant expressions from (25) and (26), solving these equations for $\alpha_\pm$ and putting into the second-order expansion for $\Omega^2$, we finally get frequencies in second-order as

$$\Omega_{2+} = \left[\frac{\omega^2}{2} + \frac{3}{8}bC^2 + \frac{1}{48}\left\{\begin{matrix} 18(32\omega^4 + 48\omega^2 bC^2 + 15b^2C^4) \\ -384(5a^2C^2 - 3abC^3) \end{matrix}\right\}^{\frac{1}{2}}\right]^{\frac{1}{2}}, \tag{36a}$$

and

$$\Omega_{2-} = \left[\frac{\omega^2}{2} + \frac{3}{8}bD^2 + \frac{1}{48}\left\{\begin{matrix} 18(32\omega^4 + 48\omega^2 bD^2 + 15b^2D^4) \\ -384(5a^2D^2 + 3abD^3) \end{matrix}\right\}^{\frac{1}{2}}\right]^{\frac{1}{2}}. \tag{36b}$$

Obviously, for $\Omega_{2\pm}$ to be real,

$$3(32\omega^4 + 48\omega^2 bC^2 + 15b^2C^4) > 64(5a^2C^2 - 3abC^3), \tag{37a}$$

and

$$3(32\omega^4 + 48\omega^2 bD^2 + 15b^2D^4) > 64(5a^2D^2 + 3abD^3). \tag{37b}$$

Otherwise, the method will not be applicable.

For a cubic AHO, the values of $\Omega_{2\pm}$ are obtained from Eqs. (36a) and (36b) by substituting b = 0. In this case, the conditions (37a) and (37b) become

$$\omega^2 > \sqrt{\frac{10}{3}}\, aC \qquad \text{and} \qquad \omega^2 > \sqrt{\frac{10}{3}}\, aD. \tag{38}$$

Since D > C > 0, both these conditions are fulfilled if $\omega > 1.351\ aD$. Clearly, this is in quite contrast with the restriction found for the implementation of the approach followed by Amore et al; see comments after Eqs. (33).

In view of the fact that we have gone up to the third-order for the previous procedure, we do so here as well and, thus, take

$$\omega_{1\pm} + \omega_{2\pm} + \omega_{3\pm} = 0. \tag{39}$$

Substituting Eqs. (25) - (27) into Eq. (39), we obtained the algebraic cubic equations

$$p_{3\pm}\alpha_\pm^3 + p_{2\pm}\alpha_\pm^2 + p_{1\pm}\alpha_\pm + p_{0\pm} = 0. \tag{40}$$

Here,

$$p_{3\pm} = 4608, \tag{41a}$$

$$p_{2+} = 9216\omega^2 - 3456bC^2, \tag{41b}$$

$$p_{1+} = 4608\omega^4 - 6912\omega^2 bC^2 + 7680a^2C^2 - 4608abC^3 + 216b^2C^4, \tag{41c}$$

$$p_{0+} = -3456\omega^4 bC^2 + 3840\omega^2 a^2C^2 - 2304\omega^2 abC^3 + 108\omega^2 b^2C^4 \\ +2560a^3C^3 + 2448ab^2C^5 - 8880a^2bC^4 - 81b^3C^6 \tag{41d}$$

$$p_{2-} = 9216\omega^2 - 3456bD^2 \tag{41e}$$

$$p_{1-} = 4608\omega^4 - 6912\omega^2 bD^2 + 7680a^2D^2 + 4608abD^3 + 216b^2D^4 \tag{41f}$$

$$p_{0-} = -3456\omega^4 bD^2 + 3840\omega^2 a^2D^2 + 2304\omega^2 abD^3 + 108\omega^2 b^2D^4 \\ -2560a^3D^3 - 2448ab^2D^5 - 8880a^2bD^4 - 81b^3D^6 \tag{41g}$$

The real roots of Eq. (40) together with $\Omega_{3\pm}^2 = \omega^2 + \alpha_\pm$ yield the frequency values up to third-order.

*3.3. Modified procedure of Sun et al*

As a follow up of the successful usage of the improved LP methods for the AHOs with symmetric potentials, Sun et al [12] suggested an approach to solve the problem of an AHO with mixed parity, wherein the difference in the amplitudes for the motions corresponding to q(t) > 0 and q(t) < 0 is taken care of in a completely different manner than has been done in the preceding two subsections. The potential used by them for elaboration was $V(q) = \frac{q^3}{3} + \frac{q^4}{4}$ and the numerical calculations for the angular frequency were performed through the Wu et al approach up to second-order and the Amore et al technique up to third-order.

With a view to implement their procedure, we rewrite Eq. (16) as

$$\Omega^2 q''(\tau) + (\omega^2 + \alpha)q(\tau) + \xi = \varepsilon[\xi + \alpha q(\tau) - aq^2(\tau) - bq^3(\tau)], \quad (42)$$

where $\xi$ is a constant to be determined. When expansions in (17) and (18) are substituted in (42), Eqs. (19) and (20) are modified by additional terms $-\xi$ and $\xi$, respectively, on the right hand side, while Eqs. (21) and (22) are unchanged. The differential equation obtained in place of Eq. (19) when solved according to the conditions in (23), gives

$$q_0(\tau) = \frac{C+D}{2} cos\tau - \frac{D-C}{2}. \quad (43)$$

In this expression, the explicit effect of the difference in the times taken to cover the distances for q(τ) > 0 and q(τ) < 0 is lost and a sort of average appears in (43) with τ = Ωt. In view of this, we have used only the Wu et al formalism up to second-order and have not gone up to third-order which means not opting for the Amore et al prescription for this treatment.

Solving the new differential equation so obtained in place of Eq. (20), with the relevant conditions given in (23), and then Eq. (21) as usual and eliminating the secular terms, we find that

$$\omega_1 = -2aB + \frac{3b}{4}(A^2 + 4B^2) - \alpha, \quad (44)$$

$$\omega_2 = \frac{B}{4}(8a - 9bA - 24bB) + \frac{B^2}{A}(2a - 3bB) + \frac{B\alpha}{A} + \frac{1}{384\omega_0}\begin{bmatrix} bA\{192A^2(a - 3bB) + 576b^2(a - bB) - 9bA^3\} \\ +64(a - 3bB)\{-5A^2(a - 3bB) - 12B^2(a - bB) - 12B\alpha\} \end{bmatrix}, \quad (45)$$

where

$$A = \frac{D+C}{2} \quad \text{and} \quad B = \frac{D-C}{2}. \quad (46)$$

As per Wu et al prescription up to second-order, we put $\omega_1 + \omega_2 = 0$ and after simplification get

$$\alpha = -\frac{1}{2}\left(\omega^2 + 2aB - \frac{3bA^2}{4} - 3bB^2\right) + \frac{1}{2}\begin{bmatrix} \left(\omega^2 + 2aB - \frac{3bA^2}{4} - 3bB^2\right)^2 \\ - \left\{4S/\left(\frac{B}{A} - 1\right)\right\} \end{bmatrix}^{\frac{1}{2}}, \quad (47)$$

with

$$S = \omega^2\left[2a\frac{B^2}{A} + \frac{3}{4}bA^2\left(1 - \frac{3B}{A}\right) - 3bB^2\left(1 + \frac{B}{A}\right)\right] - a^2\left(\frac{5}{6}A^2 + 2B^2\right) + ab\left(\frac{A^3}{2} + 5A^2B + \frac{3}{2}AB^2 + 8B^3\right) - b^2\left(\frac{3A^4}{128} + \frac{3A^3B}{2} + \frac{15A^2B^2}{2} + \frac{3}{2}AB^3 + 6B^4\right). \quad (48)$$

Finally, $\Omega^2$ equals $\omega^2 + \alpha$.

For a cubic AHO, the angular frequency up to second-order is found to be given by

$$\Omega^2 = \frac{\omega^2}{2} - aB + \frac{1}{2}\left[(\omega^2 + 2aB)^2 - 4\left(\frac{B}{A} - 1\right)^{-1}\left\{\frac{2aB^2\omega^2}{A} - a^2\left(\frac{5}{6}A^2 + 2B^2\right)\right\}\right]^{\frac{1}{2}}. \quad (49)$$

*3.4. Approximate time-periods*

Having determined $\Omega_\pm$ through the Amore et al or the Wu et al approaches, we find the times taken by the AHO to cover the path for q(t) > 0 and q(t) < 0 in one oscillation from $T_\pm = \pi/\Omega_\pm$ and, thus, the time period of the AHO as $T = T_+ + T_-$. These expressions will be also used for other techniques to be described in the later sections. Furthermore, for the Sun et al prescription and for a quartic AHO $T = 2\pi/\Omega$.

## 4. Harmonic Balance Method and its other versions

It is well known that the HBM offers a relatively simple and quite reliable tool for determining the approximate value of the frequency of an AHO and has been quite often used for this purpose. The versatility of this technique in bringing out the effects of nonlinearity in the periodic systems has been well summarized by Peng et al [27]. The essential tenet of this method is to transform the non-linear differential equation describing the motion of an AHO into a set of nonlinear algebraic equations using a truncated Fourier series of pre-decided order as approximate solution. Besides, rational functions based ansatz have been also employed [6, 7,13-18]. Thus, up to third-order of approximation, we have taken

$$q(t) = C_1 cos(\Omega_+ t) + C_2 \cos(3\Omega_+ t) + C_3 \cos(5\Omega_+ t), \quad (50a)$$

with

$$q(0) = C_1 + C_2 + C_3 = C \quad (50b)$$

for q(t) > 0. The corresponding expressions for q(t) < 0 read

$$\text{q(t)}= D_1 cos(\Omega_- t) + D_2 \cos(3\Omega_- t) + D_3 \cos(5\Omega_- t) \quad (51a)$$

and

$$q(\pi) = -(D_1 + D_2 + D_3) = -\text{D}. \quad (51b)$$

We rewrite Eq. (3) as

$$\ddot{q}(t) + \omega^2 q(t) + F\big(q(t)\big) = 0, \quad (52)$$

where

$$F\big(q(t)\big) = aq^2(t) + bq^3(t). \quad (53)$$

In order to implement the HBM, we have expressed F($q(t)$) as a Fourier series up to third harmonic as

$$F\big(q(t)\big) = \beta_1 \cos(\Omega_+ t) + \beta_2 cos(3\Omega_+ t) + \beta_3 \cos(5\Omega_+ t) \quad (54)$$

for q(t) > 0. The Fourier coefficients $\beta_l$ are given by

$$\beta_l = \frac{4}{\pi}\int_0^{\frac{\pi}{2}} F\big(q(t)\big)\cos[(2l-1)\tau_+]d\tau_+ \tag{55}$$

with $\tau_+ = \Omega_+ t$ and F(q(t)) as obtained from Eq. (53) after substituting q(t) from Eq. (50a). F(q(t)) so determined from (54) was substituted into Eq. (52) where the other two terms were obtained with the help of (50). Equating to zero the coefficients of $cos(\Omega_+ t)$, $cos(3\Omega_+ t)$ and $\cos(5\Omega_+ t)$ in the expression so found, we got the following three simultaneous nonlinear algebraic equations.

$$(\Omega^2-\omega^2)C_1 - \frac{8a}{3\pi}\left[C_1^2 + \frac{2C_1C_2}{5} - \frac{2C_1C_3}{35} + \frac{10C_2C_3}{21} + \frac{27C_2^2}{35} + \frac{25C_3^2}{33}\right] - \frac{3b}{4}[C_1^3 + C_1^2C_2 + 2C_1C_2C_3 + 2C_1C_2^2 + 2C_1C_3^2 + C_2^2C_3] = 0, \tag{56}$$

$$(9\Omega^2-\omega^2)C_2 - \frac{8a}{15\pi}\left[C_1^2 + \frac{54C_1C_2}{7} + \frac{50C_1C_3}{21} + \frac{54C_2C_3}{11} - \frac{5C_2^2}{3} - \frac{125C_3^2}{91}\right] - \frac{b}{4}[C_1^3 + 6C_1^2C_2 + 3C_1^2C_3 + 6C_1C_2C_3 + 6C_2C_3^2 + 3C_2^3] = 0, \tag{57}$$

$$(25\Omega^2-\omega^2)C_3 - \frac{8a}{105\pi}\left[-C_1^2 + \frac{50C_1C_2}{3} + \frac{1750C_1C_3}{33} - \frac{250C_2C_3}{13} + \frac{189C_2^2}{11} + 7C_3^2\right] - \frac{3b}{4}[C_1^2C_2 + 2C_1^2C_3 + C_1C_2^2 + 2C_2^2C_3 + C_3^3] = 0. \tag{58}$$

Since $C_3 = C - C_1 - C_2$, Eqs. (56) - (58) involve three unknowns, $\Omega^2$, $C_1$ and $C_2$and these have been numerically determined employing Newton-Raphson method. A similar procedure for q(t) < 0 yielded the three equations as

$$(\Omega^2-\omega^2)D_1 + \frac{8a}{3\pi}\left[D_1^2 + \frac{2D_1D_2}{5} - \frac{2D_1D_3}{35} + \frac{10D_2D_3}{21} + \frac{27D_2^2}{35} + \frac{25D_3^2}{33}\right] - \frac{3b}{4}[D_1^3 + D_1^2D_2 + 2D_1D_2D_3 + 2D_1D_2^2 + 2D_1D_3^2 + D_2^2D_3] = 0, \tag{59}$$

$$(9\Omega^2-\omega^2)D_2 + \frac{8a}{15\pi}\left[D_1^2 + \frac{54D_1D_2}{7} + \frac{50D_1D_3}{21} + \frac{54D_2D_3}{11} - \frac{5D_2^2}{3} - \frac{125D_3^2}{91}\right] - \frac{b}{4}[D_1^3 + 6D_1^2D_2 + 3D_1^2D_3 + 6D_1D_2D_3 + 6D_2D_3^2 + 3D_2^3] = 0, \tag{60}$$

$$(25\Omega^2-\omega^2)D_3 + \frac{8a}{105\pi}\left[-D_1^2 + \frac{50D_1D_2}{3} + \frac{1750D_1D_3}{33} - \frac{250D_2D_3}{13} + \frac{189D_2^2}{11} + 7D_3^2\right] - \frac{3b}{4}[D_1^2D_2 + 2D_1^2D_3 + D_1D_2^2 + 2D_2^2D_3 + D_3^3] = 0. \tag{61}$$

with $D_3 = D - D_1 - D_2$.

It may be pointed out that if the treatment is kept only up to second-order, then a set of only two algebraic equations is obtained for each of the situations q(t) > 0 and q(t) < 0. For the former, these equations read

$$(\Omega^2-\omega^2)C_1 - \frac{8a}{3\pi}\left[C_1^2 + \frac{2C_1C_2}{5} + \frac{27C_2^2}{35}\right] - \frac{3b}{4}[C_1^3 + C_1^2C_2 + 2C_1C_2^2] = 0, \tag{62}$$

$$(9\Omega^2-\omega^2)C_2 - \frac{8a}{15\pi}\left[C_1^2 + \frac{54C_1C_2}{7} - \frac{5C_2^2}{3}\right] - \frac{b}{4}[C_1^3 + 6C_1^2C_2 + 3C_2^3] = 0, \tag{63}$$

where $C_2 = C - C_1$. We have also found the corresponding equations for q(t) < 0.

The special cases of a cubic and a quartic AHO are, respectively, derived by putting b = 0; and a = 0, D = C in the preceding expressions.

*4.1. Belendez et al approach for rational function HBM*

Guided by the success of the rational function based HBM in finding highly accurate values of the frequency for the Duffing and other AHOs [16-18], we have used

$$q(t) = \frac{C(1+\delta_+)cos(\Omega_+ t)}{1+\delta_+ cos(2\Omega_+ t)} \quad (64)$$

for q(t) > 0 and

$$q(t) = \frac{D(1+\delta_-)cos(\Omega_- \mathrm{t})}{1+\delta_- cos(2\Omega_- t)} \quad (65)$$

for q(t) < 0 as the approximate solutions of Eq. (3). Belendez et al [18] had argued that the above expression for q(t) involves all the harmonics and, therefore, leads to results far better than the simple HBM. However, in practice they used only terms up to second-order in the parameter corresponding to $\delta_\pm$ because this was assumed and found to be quite small.

In view of the above comments, carrying out binomial expansion up to second-order in $\delta_+$ for q(t) in Eq. (64), we get

$$q(t) = \left(1+\frac{\delta_+}{2}\right) C\, cos(\Omega_+ t) - \frac{\delta_+}{2}\left(1+\frac{\delta_+}{2}\right) C \cos(3\Omega_+ \mathrm{t}) + \frac{\delta_+^2}{4} C \cos(5\Omega_+ t). \quad (66)$$

When compared with the second-order expression

$$q(t) = C_1 cos(\Omega_+ t) + (C - C_1)\cos(3\Omega_+ \mathrm{t}), \quad (67)$$

it has an additional term $\frac{\delta_+^2}{4} C \cos(5\Omega_+ t)$ which is small but does contribute to second-order in $\delta_+$. Accordingly, this procedure is expected to give results somewhat superior to the usual HBM up to second-order but poorer than the HBM up to third-order.

Employing q(t) in Eqs. (64) and (65) as the assumed solutions and following the procedure of HBM in Eqs. (52) - (55), we obtained the simultaneous algebraic equations for q(t) > 0 as

$$\left(1+\frac{\delta_+}{2}\right)(\Omega_+^2 - \omega^2) - \frac{8aC}{3\pi}\left(1+\frac{4}{5}\delta_+ + \frac{8}{35}\delta_+^2\right) - \frac{3bC^2}{4}\left(1+\delta_+ + \frac{1}{2}\delta_+^2\right) = 0, \quad (68)$$

$$\left(1+\frac{\delta_+}{2}\right)\delta_+(9\Omega_+^2 - \omega^2) + \frac{16aC}{15\pi}\left(1-\frac{20}{7}\delta_+ - \frac{24}{7}\delta_+^2\right) + \frac{bC^2}{2}\left(1-\frac{3}{2}\delta_+ - 3\delta_+^2\right) = 0; \quad (69)$$

and the relevant set of equations for q(t) < 0 involving $\Omega_-$, $\delta_-$ and D.

*4.2. Two new rational functions*

Inspired by the remarks that the rational functions appear like a Pade approximants of the same order [14] and when used in HBM offer better approximation for the solution of the problem [15-18], we have proposed two new rational functional forms for q(t) which contain two parameters rather than one in Eqs. (64) and (65). Their application to the problem at hand is described in the sequel.

*4.2.1 Relatively simple higher order rational function*

Consider, for q(t) > 0,

$$q(t) = C\,\frac{(1+\delta_{1+})cos(\Omega_+ t)+\delta_{2+}\cos(3\Omega_+ t)}{1+(\delta_{1+}+\delta_{2+})cos(2\Omega_+ t)}, \tag{70}$$

where both $\delta_{1+}$ and $\delta_{2+}$ are assumed to be sufficiently small. Its binomial expansion up to second-order in $\delta_{1+}$ as well as $\delta_{2+}$ yields

$$q(t) = C\left[\begin{array}{r}\left(1+\frac{\delta_{1+}}{2}-\frac{\delta_{2+}}{2}\right)\,cos(\Omega_+ t) + \left(-\frac{\delta_{1+}}{2}+\frac{\delta_{2+}}{2}-\frac{\delta_{1+}^2}{4}+\frac{\delta_{2+}^2}{4}\right)\cos(3\Omega_+ t)\\ +\left(\frac{\delta_{1+}^2}{4}-\frac{\delta_{2+}^2}{4}\right)\cos(5\Omega_+ t)\end{array}\right], \tag{71}$$

whose difference from Eq. (66) is obvious. Since the present form of q(t) involves three unknowns $\Omega_+$, $\delta_{1+}$ and $\delta_{2+}$, the use of this approximation for q(t) gave us the following three algebraic simultaneous equations for the HBM.

$$\left(1+\frac{\delta_{1+}-\delta_{2+}}{2}\right)(\Omega_+^2-\omega^2)-\frac{8aC}{3\pi}\left(1+\frac{4}{5}\delta_{1+}+\frac{8}{35}\delta_{1+}^2-\frac{24}{35}\delta_{1+}\delta_{2+}-\frac{4}{5}\delta_{2+}+\frac{16}{35}\delta_{2+}^2\right)$$
$$-\frac{3bC^2}{4}\left(1+\delta_{1+}+\frac{1}{2}\delta_{1+}^2-\frac{3}{2}\delta_{1+}\delta_{2+}-\delta_{2+}+\delta_{2+}^2\right)=0, \tag{72}$$

$$\left(\delta_{1+}+\frac{1}{2}\delta_{1+}^2-\delta_{2+}-\frac{1}{2}\delta_{2+}^2\right)(9\Omega_+^2-\omega^2)+\frac{16aC}{15\pi}\left(1-\frac{20}{7}\delta_{1+}-\frac{24}{7}\delta_{1+}^2+\frac{88}{21}\delta_{1+}\delta_{2+}+\right.$$
$$\left.\frac{20}{7}\delta_{2+}-\frac{16}{21}\delta_{2+}^2\right)+\frac{bC^2}{2}\left(1-\frac{3}{2}\delta_{1+}-3\delta_{1+}^2+\frac{9}{2}\delta_{1+}\delta_{2+}+\frac{3}{2}\delta_{2+}-\frac{3}{2}\delta_{2+}^2\right)=0, \tag{73}$$

$$(\delta_{1+}^2-\delta_{2+}^2)(25\Omega_+^2-\omega^2)+\frac{32aC}{105\pi}\left(1+\frac{28}{3}\delta_{1+}-\frac{296}{33}\delta_{1+}^2-\frac{8}{33}\delta_{1+}\delta_{2+}-\frac{28}{3}\delta_{2+}+\right.$$
$$\left.\frac{304}{33}\delta_{2+}^2\right)+\frac{3}{2}bC^2(\delta_{1+}-\delta_{1+}\delta_{2+}-\delta_{2+}+\delta_{2+}^2)=0. \tag{74}$$

We have similarly derived the set of three equations involving $\Omega_-$, $\delta_{1-}$ and $\delta_{2-}$.

Taking b=0 in the above three equations and the corresponding ones for q(t) < 0, we obtained expressions for the cubic potential AHO. Similarly, putting a=0 in the Eqs. (72) – (74), we found the simultaneous algebraic equations for the quartic potential AHO.

*4.2.2. The second higher-order rational function*

As a possible improvement over the previous suggestion, we have also tried another rational function, namely,

$$q(t) = C\,\frac{(1+\delta_{1+})cos(\Omega_+ t)+\delta_{2+}cos(3\Omega_+ t)}{1+\delta_{1+}cos(2\Omega_+ t)+\delta_{2+}\cos(4\Omega_+ t)} \tag{75}$$

for q(t) > 0. Obviously, this is more complicated than the one talked about in section 4.2.1. In this case also $\delta_{1+}$ and $\delta_{2+}$ are taken to be small so that binomial expansion of (75) up to the second-order in these parameters leads to

$$q(t) = C\left[\begin{array}{c}\left(1+\frac{\delta_{1+}}{2}\right)\,cos(\Omega_+ t)-\left(\frac{\delta_{1+}}{2}-\frac{\delta_{2+}}{2}+\frac{\delta_{1+}^2}{4}\right)\cos(3\Omega_+ t)\\ -\left(\frac{\delta_{1+}\delta_{2+}}{2}-\frac{\delta_{1+}^2}{4}+\frac{\delta_{2+}}{2}\right)\cos(5\Omega_+ t)+\left(\frac{\delta_{1+}\delta_{2+}}{2}-\frac{\delta_{2+}^2}{4}\right)\cos(7\Omega_+ t)\\ +\frac{\delta_{2+}^2}{4}\cos(9\Omega_+ t)\end{array}\right]. \tag{76}$$

The presence of higher harmonics makes the expression to be of much higher-order.

Since Eq. (75) contains three unknown parameters $\Omega_+$, $\delta_{1+}$ and $\delta_{2+}$, while implementing the HBM, we have expressed F(q(t)) as Fourier series containing terms in $cos(\Omega_+ t)$, $\cos(3\Omega_+ t)$ and $\cos(5\Omega_+ t)$. Proceeding as earlier, we obtained the following simultaneous equations

$$\left(1+\frac{\delta_{1+}}{2}\right)(\Omega_+^2-\omega^2)-\frac{8aC}{3\pi}\left(1+\frac{4}{5}\delta_{1+}+\frac{8}{35}\delta_{1+}^2-\frac{4}{35}\delta_{1+}\delta_{2+}+\frac{8}{35}\delta_{2+}+\frac{296}{1155}\delta_{2+}^2\right)$$
$$-\frac{3bC^2}{4}\left(1+\delta_{1+}+\frac{1}{2}\delta_{1+}^2+\frac{1}{2}\delta_{2+}+\frac{1}{2}\delta_{2+}^2\right)=0, \tag{77}$$

$$\left(\delta_{1+}+\frac{1}{2}\delta_{1+}^2-\delta_{2+}\right)(9\Omega_+^2-\omega^2)+\frac{16aC}{15\pi}\left(1-\frac{20}{7}\delta_{1+}-\frac{24}{7}\delta_{1+}^2+\frac{460}{231}\delta_{1+}\delta_{2+}+\right.$$
$$\left.\frac{8}{3}\delta_{2+}-\frac{5528}{3003}\delta_{2+}^2\right)+\frac{bC^2}{2}\left(1-\frac{3}{2}\delta_{1+}-3\delta_{1+}^2+\frac{3}{2}\delta_{1+}\delta_{2+}+\frac{3}{2}\delta_{2+}-\frac{3}{2}\delta_{2+}^2\right)=0, \tag{78}$$

$$(\delta_{1+}^2-2\delta_{1+}\delta_{2+}-2\delta_{2+})(25\Omega_+^2-\omega^2)+\frac{32aC}{105\pi}\left(1+\frac{28}{3}\delta_{1+}-\frac{296}{33}\delta_{1+}^2+\frac{17348}{429}\delta_{1+}\delta_{2+}+\right.$$
$$\left.\frac{200}{11}\delta_{2+}-\frac{824}{143}\delta_{2+}^2\right)+\frac{3}{2}bC^2(\delta_{1+}+3\delta_{1+}\delta_{2+}+\delta_{2+})=0. \tag{79}$$

Needless to mention that the corresponding equations for q(t) < 0 have been also determined and the statements regarding the cubic and the quartic potential AHOs at the end of section 4.2.1 hold good here as well.

## 5. Energy balance method and its modified variants

The EBM put forward by He [19] employed trial function of first-order, involved collocation of a residual function for a specific value of time and predicted angular frequency for a quartic potential AHO having ω = b = 1 with an error of 0.39 % like many other simple techniques [28-31]. However, recently Durmaz and Kaya [20] proved that the usage of expressions for q up to second- and third- order together with a different approach for handling the residual function significantly ameliorates the accuracy. The first step in the EBM as improved in [20] is to define an instantaneous value of residual R(t), which is difference between the energy expressions at that instant of time and at the turning points for that oscillatory motion. This is evaluated using the trial solution q(t) of desired approximation n and, then, the unknown quantities, Ω and the coefficients in the expression for q(t), are found by solving the simultaneous nonlinear algebraic equations obtained by setting the weighted integrals of R(t), with weightage cos[(2j-1)Ωt] where j=1,2,…,n; equal to zero.

In the case of a cubic-quartic potential AHO, for q(t) > 0, the residual is defined as

$$\mathrm{R(t)}=\left[\frac{1}{2}m\dot{q}^2(t)+\frac{1}{2}m\omega^2q^2(t)+\frac{1}{3}maq^3(t)+\frac{1}{4}mbq^4(t)\right]$$
$$-\left[\frac{1}{2}m\omega^2C^2+\frac{1}{3}maC^3+\frac{1}{4}mbC^4\right]. \tag{80}$$

To keep the order of approximation same as in the previous two sections, we have used the expression for q(t) given by Eqs. (50a) and (50b) as the trial function. Substituting these into Eq. (80), using $\Omega_+ t=\tau_+$ and asserting that

$$\int_0^{\frac{\pi}{2}} R(\tau_+)\cos[(2j-1)\,\tau_+]d\tau_+=0 \tag{81}$$

where, j=1,2 and 3; we obtained the following set of algebraic equations in $\Omega_+$, $C_1$, $C_2$, and $C_3=C-C_1-C_2$.

$$\frac{2}{3}\Omega_+^2\left[C_1^2+\frac{18}{5}C_1C_2-\frac{10}{7}C_1C_3+\frac{110}{7}C_2C_3+\frac{459}{35}C_2^2+\frac{1225}{33}C_3^2\right]$$
$$+\frac{4}{3}\omega^2\left[C_1^2+\frac{2}{5}C_1C_2-\frac{2}{35}C_1C_3+\frac{10}{21}C_2C_3+\frac{27}{35}C_2^2+\frac{25}{33}C_3^2\right]$$
$$+\frac{\pi a}{4}[C_1^3+C_1^2C_2+2C_1C_2^2+2C_1C_2C_3+2C_1C_3^2+C_2^2C_3]$$
$$+\frac{8b}{15}\left[C_1^4+\frac{12}{7}C_1^3C_2+\frac{4}{21}C_1^3C_3+\frac{26}{7}C_1^2C_2^2+\frac{348}{77}C_1^2C_2C_3+\frac{3750}{1001}C_1^2C_3^2+\frac{172}{231}C_1C_2^3+\right.$$
$$\frac{3548}{1001}C_1C_2^2C_3+\frac{1532}{1001}C_1C_2C_3^2-\frac{164}{1547}C_1C_3^3+\frac{729}{1001}C_2^4+\frac{76}{91}C_2^3C_3+\frac{3694}{1309}C_2^2C_3^2+$$
$$\left.\frac{287500}{323323}C_2C_3^3+\frac{3125}{4389}C_3^4\right]-2C^2\left[\omega^2+\frac{2}{3}aC+\frac{b}{2}C^2\right]=0, \qquad (82)$$

$$\frac{14}{15}\Omega_+^2\left[-C_1^2+\frac{54}{49}C_1C_2+\frac{850}{147}C_1C_3+\frac{1350}{77}C_2C_3-\frac{15}{7}C_2^2-\frac{5125}{637}C_3^2\right]$$
$$+\frac{4}{15}\omega^2\left[C_1^2+\frac{54}{7}C_1C_2+\frac{50}{21}C_1C_3+\frac{54}{11}C_2C_3-\frac{5}{3}C_2^2-\frac{125}{91}C_3^2\right]$$
$$+\frac{\pi a}{12}[C_1^3+6C_1^2C_2+3C_1^2C_3+6C_1C_2C_3+3C_2^3+6C_2C_3^2]$$
$$+\frac{8b}{35}\left[C_1^4+\frac{52}{9}C_1^3C_2+\frac{116}{33}C_1^3C_3+\frac{86}{33}C_1^2C_2^2+\frac{3548}{429}C_1^2C_2C_3+\frac{766}{429}C_1^2C_3^2+\right.$$
$$\frac{972}{143}C_1C_2^3+\frac{76}{13}C_1C_2^2C_3+\frac{7388}{561}C_1C_2C_3^2+\frac{287500}{138567}C_1C_3^3-\frac{7}{9}C_2^4+\frac{972}{187}C_2^3C_3-$$
$$\left.\frac{1250}{741}C_2^2C_3^2+\frac{428}{99}C_2C_3^3-\frac{3125}{5083}C_3^4\right]+\frac{2}{3}C^2\left[\omega^2+\frac{2}{3}aC+\frac{b}{2}C^2\right]=0, \qquad (83)$$

$$\frac{46}{105}\Omega_+^2\left[C_1^2-\frac{250}{23}C_1C_2+\frac{1750}{759}C_1C_3-\frac{2250}{299}C_2C_3-\frac{1323}{253}C_2^2+\frac{175}{23}C_3^2\right]$$
$$+\frac{4}{105}\omega^2\left[-C_1^2+\frac{50}{3}C_1C_2+\frac{1750}{33}C_1C_3-\frac{250}{13}C_2C_3+\frac{189}{11}C_2^2+7C_3^2\right]$$
$$+\frac{\pi a}{4}[C_1^2C_2+2C_1^2C_3+C_1C_2^2+2C_2^2C_3+C_3^3]$$
$$+\frac{8b}{315}\left[C_1^4+\frac{348}{11}C_1^3C_2+\frac{7500}{143}C_1^3C_3+\frac{5322}{143}C_1^2C_2^2+\frac{4596}{143}C_1^2C_2C_3-\frac{738}{221}C_1^2C_3^2+\right.$$
$$\frac{228}{13}C_1C_2^3+\frac{22164}{187}C_1C_2^2C_3+\frac{2587500}{46189}C_1C_2C_3^2+\frac{12500}{209}C_1C_3^3+\frac{2187}{187}C_2^4-$$
$$\left.\frac{2500}{247}C_2^3C_3+\frac{642}{11}C_2^2C_3^2-\frac{112500}{5083}C_2C_3^3+\frac{21}{5}C_3^4\right]-\frac{2}{5}C^2\left[\omega^2+\frac{2}{3}aC+\frac{b}{2}C^2\right]=0. \qquad (84)$$

Here, too similar set of simultaneous algebraic equations was derived for $q(t) < 0$.

It may be noted that putting $b = 0$ in these expressions produced results for a cubic potential AHO while $a = 0$ led to the findings for a quartic AHO. Furthermore, by taking $C_3 = 0$ in Eqs. (82) and (83) and omitting Eq. (84) we got equations corresponding to the second-order EBM. Same comments hold good for the relevant equations for $q(t) < 0$.

### *5.1. EBM with rational function*

Using the rational function in Eq. (64) as the approximate solution for $q(t) > 0$ and for the EBM, the simultaneous algebraic equations obtained as

$$\frac{\Omega_+^2}{6}\left[1-\frac{4}{5}\delta+\frac{48}{35}\delta^2\right]+\frac{\omega^2}{3}\left[1+\frac{4}{5}\delta+\frac{8}{35}\delta^2\right]+\frac{\pi aC}{16}\left[1+\delta+\frac{\delta^2}{2}\right]$$
$$+\frac{2bC^2}{15}\left[1+\frac{8}{7}\delta+\frac{16}{21}\delta^2\right]-\frac{1}{2}\left[\omega^2+\frac{2}{3}aC+\frac{1}{2}bC^2\right]=0, \qquad (85)$$

$$-\frac{7\Omega_+^2}{30}\left[1+\frac{76}{49}\delta-\frac{16}{147}\delta^2\right]+\frac{\omega^2}{15}\left[1-\frac{20}{7}\delta-\frac{24}{7}\delta^2\right]+\frac{\pi aC}{48}\left[1-\frac{3}{2}\delta-3\delta^2\right]$$
$$+\frac{2bC^2}{35}\left[1-\frac{8}{9}\delta-\frac{272}{99}\delta^2\right]+\frac{1}{6}\left[\omega^2+\frac{2}{3}aC+\frac{1}{2}bC^2\right]=0. \quad (86)$$

A similar pair of algebraic equations for q(t) < 0 and the special cases for the cubic AHO (b = 0) and the quartic AHO (a = 0) were also obtained.

### *5.2. EBM with higher-order rational function*

The EBM has also been implemented with the trial function as proposed in Eq. (75) for q(t) > 0 and the corresponding expression for q(t) < 0. For the former, this led to the following algebraic equations.

$$\frac{\Omega_+^2}{3}\left[1-\frac{4}{5}\delta_{1+}+\frac{48}{35}\delta_{1+}^2-\frac{25}{84}\delta_{1+}\delta_{2+}+\frac{88}{35}\delta_{2+}+\frac{9496}{1155}\delta_{2+}^2\right]+\frac{2\omega^2}{3}\left[1+\frac{4}{5}\delta_{1+}+\frac{8}{35}\delta_{1+}^2-\frac{4}{35}\delta_{1+}\delta_{2+}+\frac{8}{35}\delta_{2+}+\frac{296}{1155}\delta_{2+}^2\right]+\frac{\pi aC}{8}\left[1+\delta_{1+}+\frac{1}{2}\delta_{1+}^2+\frac{1}{2}\delta_{2+}+\frac{1}{2}\delta_{2+}^2\right]+\frac{4bC^2}{15}\left[1+\frac{8}{7}\delta_{1+}+\frac{16}{21}\delta_{1+}^2+\frac{24}{77}\delta_{1+}\delta_{2+}+\frac{16}{21}\delta_{2+}+\frac{2224}{3000}\delta_{2+}^2\right]-\left[\omega^2+\frac{2}{3}aC+\frac{1}{2}bC^2\right]=0, \quad (87)$$

$$-\frac{7}{15}\Omega_+^2\left[1+\frac{76}{49}\delta_{1+}-\frac{16}{147}\delta_{1+}^2-\frac{967}{2156}\delta_{1+}\delta_{2+}+\frac{344}{147}\delta_{2+}+\frac{123752}{21021}\delta_{2+}^2\right]+\frac{2}{15}\omega^2\left[1-\frac{20}{7}\delta_{1+}-\frac{24}{7}\delta_{1+}^2+\frac{360}{231}\delta_{1+}\delta_{2+}+\frac{8}{3}\delta_{2+}-\frac{5528}{3003}\delta_{2+}^2\right]+\frac{\pi aC}{24}\left[1-\frac{3}{2}\delta_{1+}-3\delta_{1+}^2+\frac{3}{2}\delta_{1+}\delta_{2+}+\frac{3}{2}\delta_{2+}-\frac{3}{2}\delta_{2+}^2\right]+\frac{4bC^2}{35}\left[1-\frac{8}{9}\delta_{1+}-\frac{272}{99}\delta_{1+}^2+\frac{136}{143}\delta_{1+}\delta_{2+}+\frac{112}{99}\delta_{2+}-\frac{1424}{1287}\delta_{2+}^2\right]+\frac{1}{3}\left[\omega^2+\frac{2}{3}aC+\frac{1}{2}bC^2\right]=0, \quad (88)$$

$$\frac{23}{105}\Omega_+^2\left[1+\frac{148}{23}\delta_{1+}+\frac{3760}{759}\delta_{1+}^2+\frac{3593}{1196}\delta_{1+}\delta_{2+}-\frac{5000}{759}\delta_{2+}-\frac{13752}{3289}\delta_{2+}^2\right]-\frac{2}{105}\omega^2\left[1+\frac{28}{3}\delta_{1+}-\frac{296}{33}\delta_{1+}^2+\frac{17348}{429}\delta_{1+}\delta_{2+}+\frac{200}{11}\delta_{2+}-\frac{824}{143}\delta_{2+}^2\right]-\frac{\pi aC}{16}\left[\delta_{1+}+3\delta_{1+}\delta_{2+}+\delta_{2+}\right]+\frac{4bC^2}{315}\left[1-\frac{152}{11}\delta_{1+}-\frac{1104}{143}\delta_{1+}^2-\frac{5240}{143}\delta_{1+}\delta_{2+}-\frac{1488}{143}\delta_{2+}-\frac{15376}{2431}\delta_{2+}^2\right]-\frac{1}{5}\left[\omega^2+\frac{2}{3}aC+\frac{1}{2}bC^2\right]=0. \quad (89)$$

Once again the set of equations involving $\Omega_-$, $\delta_{1-}$ and $\delta_{2-}$ found through the use of the corresponding trial function for q(t) < 0 were obtained. Also the relevant expressions pertaining to b = 0 and a = 0 were determined.

## 6. Fourier series technique

With a view to employing the FSE for the determination of angular frequency or time-period in terms of the parameters of a cubic - quartic potential AHO, we used equation of motion in the form given by Eqs. (52) and (53). Keeping in mind the boundary conditions projected in Eq. (23), we have assumed the approximate solution as the Fourier cosine series expansion and for q(t) > 0 have taken this as

$$q(t)=\sum_{k=1}^{\infty}C_k\cos[(2\mathrm{k}-1)\Omega_+\mathrm{t}]. \quad (90)$$

Here, $\{C_k\}$ are the unknown Fourier coefficients. Obviously, $\mathrm{q}\ (0)\equiv \mathrm{C}=\sum_{k=1}^{\infty}C_k$ . Substituting (90) into (52), multiplying with $\cos[(2j-1)\,\Omega_+ t]$, and integrating with respect to time over the interval $\left[0,\frac{\pi}{2\Omega_+}\right]$ to utilize the orthogonality property, we found

$$C_j = \frac{4}{\pi}\frac{\int_0^{\pi/2} F\big(q(\tau_+)\big)\cos[(2j-1)\tau_+]d\tau_+}{(2j-1)^2\Omega_+^2-\omega^2} \tag{91}$$

and

$$\mathrm{C} = \frac{4}{\pi}\sum_{j=1}^{\infty}\frac{\int_0^{\pi/2} F\big(q(\tau_+)\big)\cos[(2j-1)\tau_+]d\tau_+}{(2j-1)^2\Omega_+^2-\omega^2}, \tag{92}$$

with $\tau_+ = \Omega_+ t$. Clearly, determining $F\big(q(\tau_+)\big)$ is very crucial and doing so by using the complete expression in Eq. (90) makes the things extremely cumbersome. So guided by the experience of Mendez et al [21] and Delamotte [22], we have retained only two terms in the expansion (90) for this purpose. Thus, we have used $\mathrm{q}(\tau_+) = C_1\cos\tau_+ + C_2\cos(3\tau_+)$ in Eq. (52). The resulting expression, when substituted into Eqs. (91) and (92), ultimately, yielded

$$C_1 = \frac{8a}{3\pi(\Omega_+^2-\omega^2)}\left(C_1^2 + \frac{2}{5}C_1C_2 + \frac{27}{35}C_2^2\right) + \frac{3bC_1}{4(\Omega_+^2-\omega^2)}(C_1^2 + C_1C_2 + 2C_2^2), \tag{93}$$

$$C_2 = \frac{8a}{15\pi(9\Omega_+^2-\omega^2)}\left(C_1^2 + \frac{54}{7}C_1C_2 - \frac{5}{3}C_2^2\right) + \frac{b}{4(9\Omega_+^2-\omega^2)}(C_1^3 + 6C_1^2C_2 + 3C_2^3), \tag{94}$$

and

$$\mathrm{C} = \frac{4a}{\pi}\sum_{j=1}^{\infty}\frac{(-1)^j}{[(2j-1)^2\Omega_+^2-\omega^2]}\left\{\frac{2C_1^2}{(2j-3)(2j-1)(2j+1)} - \frac{12(2j-1)C_1C_2}{(2j-5)(2j-3)(2j+1)(2j+3)} + \frac{18C_2^2}{(2j-7)(2j-1)(2j+5)}\right\} +$$
$$b\left\{C_1^3\frac{7\Omega_+^2-\omega^2}{(\Omega_+^2-\omega^2)(9\Omega_+^2-\omega^2)} + 3C_1^2C_2\frac{(71\Omega_+^4-24\Omega_+^2\omega^2+\omega^4)}{(\Omega_+^2-\omega^2)(9\Omega_+^2-\omega^2)(25\Omega_+^2-\omega^2)} +\right.$$
$$\left.3C_1C_2^2\frac{(631\Omega_+^4-56\Omega_+^2\omega^2+\omega^4)}{(\Omega_+^2-\omega^2)(25\Omega_+^2-\omega^2)(49\Omega_+^2-\omega^2)} + C_2^3\frac{63\Omega_+^2-\omega^2}{(9\Omega_+^2-\omega^2)(81\Omega_+^2-\omega^2)}\right\}. \tag{95}$$

It is pertinent to bring out the fact that Eqs. (93) and (94) are identical to the two expressions obtained in the second-order HBM, namely, Eqs. (62) and (63), respectively. However, the second-order HBM lacks an expression corresponding to Eq. (95). Therefore, it is expected that the FSE presented in this work will certainly yield results superior to the HBM of second-order, and, in view of the presence of the third relationship, Eq. (95), these should be comparable with or even better than the HBM of third-order described in section 4.

On being solved with the Newton-Raphson method, the above three simultaneous nonlinear algebraic equations gave us the numerical values of $\Omega_+$, $C_1$ and $C_2$ for a particular set of a, b and C values. Proceeding in a similar manner for the case q(t) < 0, using

$$q(t) = \sum_{l=1}^{\infty} D_l\cos[(2\mathrm{l}-1)\Omega_-\mathrm{t}], \tag{96}$$

where $q\left(\frac{\pi}{\Omega_-}\right) = -D = -\sum_{l=1}^{\infty} D_l$ and performing integration over the time interval $\left[\frac{\pi}{2\Omega_-}, \frac{\pi}{\Omega_-}\right]$, we obtained the corresponding expressions for $D_1$, $D_2$ and D. These were, in turn, employed to find $\Omega_-$, $D_1$ and $D_2$.

As usual b=0 in the above expressions produced results for a cubic potential AHO, while a = 0 with proper consideration for the associated symmetry led to the findings for a quartic AHO. Furthermore, keeping only k = 1 term in Eq. (90) and, thus, taking $C_2 = 0$, Eq. (93) is reduced to

$$C_1 = \frac{8aC_1^2}{3\pi(\Omega_+^2-\omega^2)} + \frac{3bC_1^3}{4(\Omega_+^2-\omega^2)}, \tag{97}$$

which yields the same expression for $\Omega_+$ as is obtained with HBM of the first- order applied to the cubic - quartic potential AHO.

## 7. Numerical results and discussion

The analytical expressions derived through different approaches in the preceding four sections provide approximate relationship between the angular frequency $\Omega_+$ and $\Omega_-$ for q(t) > 0 and q(t) < 0, respectively, and the cubic - quartic potential AHO parameters ω, a, b and C. These expressions have been used to determine the numerical values of $\Omega_+$ and $\Omega_-$, and thereby of the corresponding approximate periods $T_+$, $T_-$ and T, as mentioned in section 3.4, for a wide range of the values of ω, a, b and C. In order to assess the accuracy of the techniques employed and their relative strengths and weaknesses, these results have been collated with each other and also the corresponding exact values of $T_+$, $T_-$ and T determined through precise numerical integration; Eqs. (12) – (14). The representative results for some typical values of the parameters have been listed for the cubic - quartic, cubic (b = 0) and quartic (a = 0) AHOs in Tables 1 - 7. The numerals indicating the various methods dealt with in this work are the same in all the cases and have been spelled out in the caption of Table 1. The percentage difference (PD) of the outcome of a particular method for a specific set of values of ω, a, b and C from the relevant exact value (namely, [{approximate time - exact time}/ exact time} x 100]) has been included in the parenthesis after the concerned finding of the approximate approach. Obviously, PD gives a measure of the accuracy of the technique under consideration. It may be pointed out that the LPM based techniques could be used only when the values of the parameters satisfied the relevant conditions brought out in section 3. Similarly, the applicability of the approaches using higher-order rational functions suggested in the present work demands that $\delta_{1\pm}$ and $\delta_{2\pm}$ be small, a condition which is not met with while solving the corresponding simultaneous equations for some sets of the AHO parameters. The entries for such cases have been omitted in Tables 1 - 6. Furthermore, in the case of a quartic AHO, the Sun et al prescription for the LPM reduces to the second-order LPM Wu et al approach because D = C and, therefore, Table 7 does not have entries with numeral (v). It may be also mentioned that Belendez et al [18] had used approximation in their evaluation of the time period for a quartic AHO with rational function HBM, while we have solved the simultaneous equations exactly making our results slightly different from theirs.

It may be pointed out that the cases ω = 2.0, a = 5.0, b = 1.0 and C = 0.45 in Table 2 pertain to the situation $a^2 > 4\omega^2 b$ so that the corresponding D values have been obtained from the inequality in Eq.(5). Here, C = 0.45 is the largest two significant figure value for which the orbit is homoclinic. All other values of the parameters ω, a, b and C given in Tables 1 - 4, are in accord with the inequality $a^2 < 4\omega^2 b$. In a similar manner, the values of the parameters considered for the cubic AHO in Tables 5 and 6 are as per the inequality $aC < \omega^2/2$ and the cases ω = 1.0, a = 0.5, C = 0.99, and ω = 1.0, a = 1.0, C = 0.49 in Table 5 are the two extreme examples of the homoclinic orbits for the listed ω and a values as the higher values of C make the orbits heteroclinic.

**Table 1**

A comparison of the time periods for q(t) > 0, q(t) < 0 and total in respect of cubic - quartic potential AHO with ω = 1.0 and different values of a, b and C, found through (i) Exact integration, (ii) LPM Amore et al approach, (iii) LPM Wu et al prescription (third-order), (iv) LPM Wu et al prescription (second-order), (v) LPM Sun et al prescription (second-order), (vi) HBM (third-order), (vii) HBM (second-order), (viii) HBM (Rational function approach), (ix) HBM (New first rational function approach), (x) HBM (New second rational function approach), (xi) EBM (third-order), (xii) EBM (second-order), (xiii) EBM (Rational function approach), (xiv) EBM (New second rational function approach), (xv) FSE.

| | $T_+$ | $T_-$ | T |
|---|---|---|---|
| **a=0.5 b=0.5 C=0.5 D= 0.58744703** | | | |
| (i) | 2.75064418 | 3.34880986 | 6.09945404 |
| (ii) | 3.05022937(11.09) | 3.04641115(-9.03) | 6.09664052(-0.046) |
| (iii) | 3.05009363(10.89) | 3.03259387(-9.44) | 6.08268750(-0.27) |
| (iv) | 3.05285907(11.10) | 3.08836351(-7.78) | 6.14122258(0.68) |
| (v) | | | 6.10293571(0.057) |
| (vi) | 2.75063447(-3.53x10$^{-4}$) | 3.34881130(4.30x10$^{-5}$) | 6.09944576(-1.36x10$^{-4}$) |
| (vii) | 2.75069845((1.97x10$^{-3}$) | 3.34881419(1.29x10$^{-4}$) | 6.09951263((9.61x10$^{-4}$) |
| (viii) | 2.75072030(2.77x10$^{-3}$) | 3.34881418(1.29x10$^{-4}$) | 6.09953448(1.32x10$^{-3}$) |
| (ix) | 2.75063456(-3.50x10$^{-4}$) | 3.34881124(4.12x10$^{-5}$) | 6.09944579(-1.35x10$^{-4}$) |
| (x) | 2.75063320(-3.99x10$^{-4}$) | 3.34881128(4.24x10$^{-5}$) | 6.09944449(-1.57x10$^{-4}$) |
| (xi) | 2.75070736(2.30x10$^{-3}$) | 3.34860036(-6.55x10$^{-3}$) | 6.09930772(-2.40x10$^{-3}$) |
| (xii) | 2.74944917(-4.34x10$^{-2}$) | 3.35236689(0.11) | 6.10181606(0.039) |
| (xiii) | 2.74897235(-6.08x10$^{-2}$) | 3.35235866(0.11) | 6.10133101(0.031) |
| (xiv) | 2.75072564(2.96x10$^{-3}$) | 3.34860355(-6.16x10$^{-3}$) | 6.09932919(-2.05x10$^{-3}$) |
| (xv) | 2.75064756(1.23x10$^{-4}$) | 3.34881959(2.91x10$^{-4}$) | 6.09946715(2.15x10$^{-4}$) |
| **a=1.0 b=1.0 C=1.0 D=1.42574609** | | | |
| (i) | 1.95819819 | 2.76320302 | 4.72140121 |
| (ii) | 2.44928720(25.08) | 2.29601903(-16.91) | 4.74530623(0.51) |
| (iii) | 2.38101923(21.59) | 1.98552490(-28.14) | 4.36654413(-7.52) |
| (v) | | | 4.74103626(0.42) |
| (vi) | 1.95815911(-2.00x10$^{-3}$) | 2.76333484(-4.77x10$^{-3}$) | 4.72149394(1.96x10$^{-3}$) |
| (vii) | 1.95815863(-2.02x10$^{-3}$) | 2.76004244(-0.11) | 4.71820106(-6.78x10$^{-2}$) |
| (viii) | 1.95854365(1.76x10$^{-2}$) | 2.76118805(-7.29x10$^{-2}$) | 4.71973170(-3.54x10$^{-2}$) |
| (ix) | 1.95815367(-2.27x10$^{-3}$) | 2.76331226(3.96x10$^{-3}$) | 4.72146593(1.37x10$^{-3}$) |
| (x) | 1.95813605(-3.17x10$^{-3}$) | 2.76340455(7.29x10$^{-3}$) | 4.72154060(2.95x10$^{-3}$) |
| (xi) | 1.95823419(1.84x10$^{-3}$) | 2.76342565(8.06x10$^{-3}$) | 4.72165984(5.48x10$^{-3}$) |
| (xii) | 1.95846315(1.35x10$^{-2}$) | 2.77881176(0.56) | 4.73727491(0.34) |
| (xiii) | 1.95576589(-0.12) | 2.77265997(0.34) | 4.72842586(0.15) |
| (xiv) | 1.95821656(9.38x10$^{-4}$) | 2.76260471(-2.17x10$^{-2}$) | 4.72082127(-1.23 x10$^{-2}$) |
| (xv) | 1.95819658(-8.22x10$^{-5}$) | 2.76317900((-8.69x10$^{-4}$) | 4.72137558((-5.43x10$^{-4}$) |
| **a=1.0 b=1.0 C=1000 D=1000.66666615** | | | |
| (i) | 0.00370597 | 0.00370785 | 0.00741382 |
| (ii) | 0.00370462(-3.64x10$^{-2}$) | 0.00370415(-9.98x10$^{-2}$) | 0.00740877(-6.81x10$^{-2}$) |
| (iii) | 0.00370453(-3.89x10$^{-2}$) | 0.00370392(-0.11) | 0.00740845(-7.24x10$^{-2}$) |
| (iv) | 0.00370741(3.89x10$^{-2}$) | 0.00370871(2.32x10$^{-2}$) | 0.00741612(3.10x10$^{-2}$) |
| (v) | - | - | 0.00741617(3.17x10$^{-2}$) |
| (vi) | 0.00370564(-8.90x10$^{-3}$) | 0.00370753(-8.63x10$^{-3}$) | 0.00741317(-8.77x10$^{-3}$) |
| (vii) | 0.00369873(-0.20) | 0.00370058(-0.20) | 0.00739931(-0.20) |
| (viii) | 0.00370581(-4.32x10$^{-3}$) | 0.00370768(-4.58x10$^{-3}$) | 0.00741348(-4.59x10$^{-3}$) |
| (ix) | 0.00370522(-2.02x10$^{-2}$) | 0.00370710(-2.02x10$^{-2}$) | 0.00741232(-2.02x10$^{-2}$) |
| (x) | 0.00370516(-2.19x10$^{-2}$) | 0.00370705(-2.16x10$^{-2}$) | 0.00741221(-2.17x10$^{-2}$) |
| (xi) | 0.00370627(8.10x10$^{-3}$) | 0.00370816(8.36x10$^{-3}$) | 0.00741443(8.23x10$^{-3}$) |
| (xii) | 0.00372808(0.60) | 0.00373006(0.60) | 0.00745814(0.60) |
| (xv) | 0.00370518(-2.13x10$^{-2}$) | 0.00370706(-2.13x10$^{-2}$) | 0.00741224(-2.13x10$^{-2}$) |

**Table 2**

The time periods for a cubic - quartic AHO with $\omega = 2.0$. Labels (i) to (xv) have the same meaning as in Table 1.

| | $T_+$ | $T_-$ | T |
|---|---|---|---|
| **a=1.0 b=1.0 C=1.0 D=1.15167393** | | | |
| (i) | 1.32906741 | 1.56781316 | 2.89688056 |
| (ii) | 1.44998912(9.10) | 1.44695551(-7.71) | 2.89694463(2.21x$10^{-3}$) |
| (iii) | 1.44886179(9.01) | 1.43913459(-8.21) | 2.88799639(-0.31) |
| (iv) | 1.45318581(9.34) | 1.46838067(-6.34) | 2.92156648(0.85) |
| (v) | | | 2.89818247(4.49x$10^{-2}$) |
| (vi) | 1.32906147(-4.47x$10^{-4}$) | 1.56782039(4.61x$10^{-4}$) | 2.89688186(4.49x$10^{-5}$) |
| (vii) | 1.32908915(1.64x$10^{-3}$) | 1.56776292(-3.20x$10^{-3}$) | 2.89685208(-9.83x$10^{-4}$) |
| (viii) | 1.32911448(3.54x$10^{-3}$) | 1.56776621(-2.99x$10^{-3}$) | 2.89688069(4.49x$10^{-6}$) |
| (ix) | 1.32906139(-4.53x$10^{-4}$) | 1.56782054(4.71x$10^{-4}$) | 2.89688193(4.73x$10^{-5}$) |
| (x) | 1.32906030(-5.35x$10^{-4}$) | 1.56782090(4.94x$10^{-4}$) | 2.89688120(2.21x$10^{-5}$) |
| (xi) | 1.32909214(1.86x$10^{-3}$) | 1.56775346(-3.81x$10^{-3}$) | 2.89684560(-1.21x$10^{-3}$) |
| (xii) | 1.32871819(-2.63x$10^{-2}$) | 1.56934044(9.74x$10^{-2}$) | 2.89805863(4.07x$10^{-2}$) |
| (xiii) | 1.32831252(-5.68x$10^{-2}$) | 1.56923370(9.06x$10^{-2}$) | 2.89754622(2.30x$10^{-2}$) |
| (xiv) | 1.32909896(2.37x$10^{-3}$) | 1.56773638(-4.90x$10^{-3}$) | 2.89683534(-1.56 x$10^{-3}$) |
| (xv) | 1.32906873(9.93x$10^{-5}$) | 1.56781201(-7.34x$10^{-5}$) | 2.89688074(6.21x$10^{-6}$) |
| **a=5.0 b=1.0 C=0.45 D=0.89339566** | | | |
| (i) | 1.27715110 | 3.83120808 | 5.10835918 |
| (ii) | 1.88619540(47.69) | 2.02186011(-47.23) | 3.90805550(-23.50) |
| (iii) | 1.88531137(47.62) | 1.50681602(-60.67) | 3.39212738(-33.60) |
| (vi) | 1.27714102(-7.89x$10^{-4}$) | 3.81305325(-0.47) | 5.09019427(-0.36) |
| (vii) | 1.27721592(5.08x$10^{-3}$) | 3.94677977(3.02) | 5.22399569(2.26) |
| (viii) | 1.27723591(6.64x$10^{-3}$) | 3.86928597(0.99) | 5.14652188(0.75) |
| (ix) | 1.27714137(-7.62x$10^{-3}$) | 3.80931859(-0.57) | 5.08645996(-0.43) |
| (x) | 1.27713936(-9.19x$10^{-4}$) | 3.81982366(-0.30) | 5.09696302(-0.22) |
| (xi) | 1.27720105(3.91x$10^{-3}$) | 3.76972033(-1.60) | 5.04692138(-1.20) |
| (xii) | 1.2760(-9.01x$10^{-2}$) | 7.2045(88.05) | 8.4805(66.01) |
| (xiii) | 1.27564449(-0.12) | 3.99680378(4.32) | 5.27244827(3.21) |
| (xiv) | 1.27722253(5.59x$10^{-3}$) | 3.84662354(0.40) | 5.12384607(0.30) |
| (xv) | 1.27715701(4.63x$10^{-4}$) | 3.78905466(-1.10) | 5.06621167(-0.83) |
| **a=5.0 b=100.0 C=1.0 D=1.03210629** | | | |
| **a=0.5 b=1.0 C=10.0 D=10.32106290** | | | |
| (i) | 0.35089448 | 0.35968717 | 0.71058164 |
| (ii) | 0.35344239(0.73) | 0.35430168(-1.50) | 0.70774407(-0.40) |
| (iii) | 0.35531230(1.26) | 0.35287284(-1.89) | 0.70818514(-0.34) |
| (iv) | 0.35360449(0.77) | 0.35909841(-0.16) | 0.71270290(0.30) |
| (v) | | | 0.71100206(5.93x$10^{-2}$) |
| (vi) | 0.35087110(-6.66x$10^{-3}$) | 0.35966009(-7.53x$10^{-3}$) | 0.71053119(-7.10x$10^{-3}$) |
| (vii) | 0.35037125(-0.15) | 0.35901944(-0.19) | 0.70939069(-0.17) |
| (viii) | 0.35090642(3.40x$10^{-3}$) | 0.35964607(-1.14x$10^{-2}$) | 0.71055249(-4.10x$10^{-3}$) |
| (ix) | 0.35084233(-1.49x$10^{-2}$) | 0.35962308(-1.78x$10^{-2}$) | 0.71046541(-1.64x$10^{-2}$) |
| (x) | 0.35083696(-1.64x$10^{-2}$) | 0.35962110(-1.84x$10^{-2}$) | 0.71045806(-1.74x$10^{-2}$) |
| (xi) | 0.35091798(6.70x$10^{-3}$) | 0.35971815(8.61x$10^{-3}$) | 0.71063613(7.67x$10^{-3}$) |
| (xii) | 0.35262667(0.49) | 0.36178818(0.58) | 0.71441485(0.54) |
| (xiii) | 0.35082548(-1.97x$10^{-2}$) | 0.35977106(2.33x$10^{-2}$) | 0.71059653(2.10x$10^{-3}$) |
| (xiv) | 0.35080366(-2.59x$10^{-2}$) | 0.35957760(-3.05x$10^{-2}$) | 0.71038126(-2.82x$10^{-2}$) |
| (xv) | 0.35083949(-1.57x$10^{-2}$) | 0.35961780(-1.93x$10^{-2}$) | 0.71045729(-1.75x$10^{-2}$) |

**Table 3**

The time periods for cubic - quartic potential AHO with ω = 5.0 and different values of a, b and C.

| | $T_+$ | $T_-$ | T |
|---|---|---|---|
| **a=1.0 b=1.0 C=1.0 D=1.02630632** | | | |
| (i) | 0.60916968 | 0.62934292 | 1.23851260 |
| (ii) | 0.61925721(1.66) | 0.61925540(-1.60) | 1.23851260(7x10$^{-7}$) |
| (iii) | 0.61925655(1.65) | 0.61925297(-1.59) | 1.23850951(-2.49x10$^{-4}$) |
| (iv) | 0.61926736(1.66) | 0.61929482(-1.60) | 1.23856218(4.00x10$^{-3}$) |
| (v) | | | 1.23851670(3.31x10$^{-4}$) |
| (vi) | 0.60916956(-1.97x10$^{-5}$) | 0.62934298(9.53x10$^{-6}$) | 1.23851254(-4.84x10$^{-6}$) |
| (vii) | 0.60917050(1.35x10$^{-4}$) | 0.62934263(-4.61x10$^{-5}$) | 1.23851313(4.28x10$^{-5}$) |
| (viii) | 0.60917060(1.51x10$^{-4}$) | 0.62934263(-4.61x10$^{-5}$) | 1.23851324(5.85x10$^{-5}$) |
| (ix) | 0.60916956(-1.97x10$^{-5}$) | 0.62934298(9.53x10$^{-6}$) | 1.23851254(5.17x10$^{-5}$) |
| (x) | 0.60916955(-1.97x10$^{-5}$) | 0.62934298(9.53x10$^{-6}$) | 1.23851253(-5.65x10$^{-6}$) |
| (xi) | 0.60917346(6.21x10$^{-4}$) | 0.62933829(-7.36x10$^{-4}$) | 1.23851175(-6.86x10$^{-5}$) |
| (xii) | 0.60910544(-1.05x10$^{-2}$) | 0.62942581(1.32x10$^{-2}$) | 1.23853125(1.51x10$^{-3}$) |
| (xiii) | 0.60909694(-1.19x10$^{-2}$) | 0.62942531(1.31x10$^{-2}$) | 1.23852225(7.79x10$^{-4}$) |
| (xiv) | 0.60917376(6.70x10$^{-4}$) | 0.62933818(-7.53x10$^{-4}$) | 1.23851194(-5.33x10$^{-5}$) |
| (xv) | 0.60916971(4.92x10$^{-6}$) | 0.62934297(7.94x10$^{-6}$) | 1.23851268(6.46x10$^{-6}$) |

**Table 4**

The time periods $T_+$, $T_-$, T for q(t) > 0, q(t) < 0 and total, respectively, for a cubic - quartic AHO having ω = 10.0.

| | $T_+$ | $T_-$ | T |
|---|---|---|---|
| **a=5.0 b=100 C=100 D=100.03333000** | | | |
| (i) | 0.00370679 | 0.00370773 | 0.00741452 |
| (ii) | 0.00370485(-5.23x10$^{-2}$) | 0.00370462(-8.39x10$^{-2}$) | 0.00740947(-6.81x10$^{-2}$) |
| (iii) | 0.00370473(-5.56x10$^{-2}$) | 0.00370443(-8.90x10$^{-2}$) | 0.00740916(-7.23x10$^{-2}$) |
| (iv) | 0.00370808(3.48x10$^{-2}$) | 0.00370873(2.70x10$^{-2}$) | 0.00741682(3.10x10$^{-2}$) |
| (v) | - | - | 0.00741684(3.13x10$^{-2}$) |
| (vi) | 0.00370646(-8.90x10$^{-3}$) | 0.00370741(-8.63x10$^{-3}$) | 0.00741387(-8.77x10$^{-3}$) |
| (vii) | 0.00369955(-0.20) | 0.00370047(-0.20) | 0.00740002(-0.20) |
| (viii) | 0.00370662(-4.59x10$^{-3}$) | 0.00370756(-4.59x10$^{-3}$) | 0.00741418(-4.59x10$^{-3}$) |
| (ix) | 0.00370604(-2.02x10$^{-2}$) | 0.00370698(-2.02x10$^{-2}$) | 0.00741302(-2.02x10$^{-2}$) |
| (x) | 0.00370598(-2.19x10$^{-2}$) | 0.00370693(-2.16x10$^{-2}$) | 0.00741291(-2.17x10$^{-2}$) |
| (xi) | 0.00370709(8.09x10$^{-3}$) | 0.00370804(8.36x10$^{-3}$) | 0.00741513(8.23x10$^{-3}$) |
| (xii) | 0.00372892(0.60) | 0.00372991(0.60) | 0.00745883(0.60) |
| (xv) | 0.00370600(-2.13x10$^{-2}$) | 0.00370694(-2.13x10$^{-2}$) | 0.00741294(-2.13x10$^{-2}$) |

**Table 5**

The time periods for an AHO governed by cubic potential and having ω = 1.0.

| | $T_+$ | $T_-$ | T |
|---|---|---|---|
| **a= 0.5 C=0.99 D=1.82201156** | | | |
| (i) | 2.63782541 | 7.24617302 | 9.88399843 |
| (ii) | 3.76774676(42.84) | 4.16217391(-42.56) | 7.92992067(-19.77) |
| (iii) | 3.76662396(42.79) | 3.36145856(-53.61) | 7.12808253(-27.88) |
| (vi) | 2.63780933(-6.10x10$^{-4}$) | 7.21432397(-0.44) | 9.85213330(-0.32) |
| (vii) | 2.63793559(4.18x10$^{-3}$) | 7.47175685(3.11) | 10.10969244(2.28) |
| (viii) | 2.63795870(5.05x10$^{-3}$) | 7.30541341(0.82) | 9.94337211(0.60) |
| (ix) | 2.63780991(-5.88x10$^{-4}$) | 7.20837148(-0.52) | 9.84618140(-0.38) |
| (x) | 2.63780726(-6.88x10$^{-4}$) | 7.22462759(-0.30) | 9.86243485(-0.22) |
| (xi) | 2.63793387(4.11x10$^{-3}$) | 7.13831053(-1.49) | 9.77624440(-1.09) |
| (xiii) | 2.63490455(-0.11) | 7.51609139(3.72) | 10.15099595(2.70) |
| (xiv) | 2.63797147(5.54x10$^{-3}$) | 7.28422868(0.53) | 9.92220015(0.39) |
| (xv) | 2.63783579(3.94x10$^{-4}$) | 7.16759445(-1.08) | 9.80543024(-0.79) |

| a=1.0 C=0.4 D=0.57250828 | | | |
|---|---|---|---|
| (i) | 2.71561804 | 4.40894847 | 7.12456652 |
| (ii) | 3.47316136(27.90) | 3.50733990(-20.45) | 6.98050126(-2.02) |
| (iii) | 3.46883885(27.74) | 3.36720009(-23.63) | 6.83603894(-4.05) |
| (v) | - | - | 7.18437512(0.84) |
| (vi) | 2.71560574(-4.53x10$^{-4}$) | 4.40852012(-9.72x10$^{-3}$) | 7.12412586(-6.19x10$^{-3}$) |
| (vii) | 2.71570319(3.14x10$^{-3}$) | 4.41215628(7.28x10$^{-2}$) | 7.12785947(4.62x10$^{-2}$) |
| (viii) | 2.71571809(3.68x10$^{-3}$) | 4.41094502(4.53x10$^{-2}$) | 7.12666311(2.94x10$^{-2}$) |
| (ix) | 2.71560612(-4.39x10$^{-4}$) | 4.40847670(-1.07x10$^{-2}$) | 7.12408282(-6.79x10$^{-3}$) |
| (x) | 2.71560441(-5.02x10$^{-4}$) | 4.40860660(-7.75x10$^{-3}$) | 7.12421101(-4.99x10$^{-3}$) |
| (xi) | 2.71572085(3.79x10$^{-3}$) | 4.40642466(-5.72x10$^{-2}$) | 7.12214551(-3.40x10$^{-2}$) |
| (xii) | 2.71341769(-8.10x10$^{-2}$) | 4.43606363(0.62) | 7.14948132(0.35) |
| (xiii) | 2.71304255(-9.48x10$^{-2}$) | 4.42558753(0.38) | 7.13863008(0.20) |
| (xiv) | 2.71575039(4.87x10$^{-3}$) | 4.40802077(-2.10x10$^{-2}$) | 7.12377116(-1.12x10$^{-2}$) |
| (xv) | 2.71562573(2.83x10$^{-4}$) | 4.40884313(-2.39x10$^{-3}$) | 7.12446886(-1.37x10$^{-3}$) |
| **a=1.0 C=0.49 D=0.87283208** | | | |
| (i) | 2.64175268 | 6.56624588 | 9.20799856 |
| (ii) | 3.74730997(41.85) | 4.06029004(-38.16) | 7.80760002(-15.21) |
| (iii) | 3.74586446(41.80) | 3.37027641(-48.67) | 7.11614087(-22.72) |
| (vi) | 2.64173678(-6.02x10$^{-4}$) | 6.55187894(-0.22) | 9.19361572(-0.16) |
| (vii) | 2.64186157(4.12x10$^{-3}$) | 6.66659984(1.53) | 9.30846141(1.09) |
| (viii) | 2.64188422(4.98x10$^{-3}$) | 6.59778892(0.48) | 9.23967313(0.34) |
| (ix) | 2.64173736(-5.80x10$^{-4}$) | 6.54917503(-0.26) | 9.19091239(-0.19) |
| (x) | 2.64173475(-6.79x10$^{-4}$) | 6.55621083(-0.15) | 9.19794558(-0.11) |
| (xi) | 2.64186091(4.10x10$^{-3}$) | 6.51300439(-0.81) | 9.15486530(-0.58) |
| (xiii) | 2.63884838(-0.11) | 6.70209439(2.07) | 9.34094276(1.44) |
| (xiv) | 2.64189810(5.50x10$^{-3}$) | 6.57902636(0.19) | 9.22092446(0.14) |
| (xv) | 2.64176291(3.87x10$^{-4}$) | 6.53801570(-0.43) | 9.17977861(-0.31) |

**Table 6**

The time periods for a cubic potential AHO with ω≠ 1.0.

| | $T_+$ | $T_-$ | T |
|---|---|---|---|
| **ω=2.0 a=1.0 C=1.0 D=1.20871215** | | | |
| (i) | 1.42700146 | 1.82320548 | 3.25020694 |
| (ii) | 1.62231147(13.69) | 1.62303371(-10.98) | 3.24534518(-0.15) |
| (iii) | 1.62163534(13.64) | 1.61886649(-11.21) | 3.24050184(-0.30) |
| (iv) | 1.61596499(13.24) | 1.64032571(-10.03) | 3.25629070(0.19) |
| (v) | - | - | 3.25282002(8.04x10$^{-2}$) |
| (vi) | 1.42699830(-2.21x10$^{-4}$) | 1.82318667(-1.03x10$^{-3}$) | 3.25018497(-6.76x10$^{-4}$) |
| (vii) | 1.42702367 (1.56x10$^{-3}$) | 1.82334484(7.64x10$^{-3}$) | 3.25036851(4.97x10$^{-3}$) |
| (viii) | 1.42702624(1.74x10$^{-3}$) | 1.82332051(6.31x10$^{-3}$) | 3.25034675(4.30x10$^{-3}$) |
| (ix) | 1.42699837(-2.17x10$^{-4}$) | 1.82318591(-1.07x10$^{-3}$) | 3.25018427(-6.97x10$^{-4}$) |
| (x) | 1.42699807(-2.38x10$^{-4}$) | 1.82318859(-9.26x10$^{-4}$) | 3.25018666(-6.24x10$^{-4}$) |
| (xi) | 1.42704376(2.96x10$^{-3}$) | 1.82299659(-1.15x10$^{-2}$) | 3.25004035(-5.13x10$^{-3}$) |
| (xii) | 1.42616163(-5.89x10$^{-2}$) | 1.82609576(0.16) | 3.25225739(6.31x10$^{-2}$) |
| (xiii) | 1.42606685(-6.55x10$^{-2}$) | 1.82559030(0.13) | 3.25165715(4.46x10$^{-2}$) |
| (xiv) | 1.42705170(3.52x10$^{-3}$) | 1.82305582(-8.21x10$^{-3}$) | 3.25010752(-3.06x10$^{-3}$) |
| (xv) | 1.42700330(1.29x10$^{-4}$) | 1.82320982(2.38x10$^{-4}$) | 3.25021312(1.90x10$^{-4}$) |
| **ω=5.0 a=1.0 C=10.0 D=14.31270696** | | | |
| (i) | 0.54312361 | 0.88178970 | 1.42491331 |
| (ii) | 0.69463227(27.90) | 0.70146798(-20.45) | 1.39610025(-2.02) |
| (iii) | 0.69376777(27.74) | 0.67344002(-23.63) | 1.36720779(-4.05) |
| (v) | - | - | 1.43687502(0.84) |
| (vi) | 0.54312115 (-4.53x10$^{-4}$) | 0.88170402(-9.72x10$^{-3}$) | 1.42482517(-6.19x10$^{-3}$) |
| (vii) | 0.54314064(3.14x$10^{-3}$) | 0.88243126(7.28x$10^{-2}$) | 1.42557189(4.62x$10^{-2}$) |
| (viii) | 0.54314362(3.68x$10^{-3}$) | 1.14252021(29.57) | 1.68566383(18.30) |
| (ix) | 0.54312122(-4.40x$10^{-4}$) | 0.88169534(-1.07x$10^{-2}$) | 1.42481656(-6.79x$10^{-3}$) |
| (x) | 0.54312088(-5.03x$10^{-4}$) | 0.88172132(-7.75x$10^{-3}$) | 1.42484220(-4.99x$10^{-3}$) |
| (xi) | 0.54314417(3.79x$10^{-3}$) | 0.88128493(-5.72x$10^{-2}$) | 1.42442910(-3.40x$10^{-2}$) |
| (xii) | 0.54268354(-8.10x$10^{-2}$) | 0.88721273(0.62) | 1.42989626(0.35) |
| (xv) | 0.54312515(2.84x$10^{-4}$) | 0.88176863(-2.39x$10^{-3}$) | 1.42489377(-1.37x$10^{-3}$) |
| **ω=10.0 a=0.5 C=10.0 D=10.34523859**<br>**ω=10.0 a=1.0 C=5.0 D=5.17261929**<br>**ω=10.0 a=5.0 C=1.0 D=1.03452386** | | | |
| (i) | 0.30770111 | 0.32129655 | 0.62899766 |
| (ii) | 0.31449835(2.21) | 0.31449837(-2.12) | 0.62899672(-1.49x$10^{-4}$) |
| (iii) | 0.31449802(2.21) | 0.31449792(-2.12) | 0.62899593(-2.75x$10^{-4}$) |
| (iv) | 0.31448771(2.20) | 0.31451087(-2.11) | 0.62899859(1.48x$10^{-4}$) |
| (v) | - | - | 0.62900003(3.77x$10^{-4}$) |
| (vi) | 0.30770108(-9.75x$10^{-6}$) | 0.32129650(-1.56x$10^{-5}$) | 0.62899757(-1.43x$10^{-5}$) |
| (vii) | 0.30770139(9.10x$10^{-5}$) | 0.32129693(1.18x$10^{-4}$) | 0.62899832(1.05x$10^{-4}$) |
| (viii) | 0.30770140(9.42x$10^{-5}$) | 0.32129692(1.15x$10^{-4}$) | 0.62899832(1.05x$10^{-4}$) |
| (ix) | 0.30770108(-9.75x$10^{-6}$) | 0.32129650(-1.56x$10^{-5}$) | 0.62899757 (-1.43x$10^{-5}$) |
| (x) | 0.30770108(-9.75x$10^{-6}$) | 0.32129650(-1.56x$10^{-5}$) | 0.62899757 (-1.43x$10^{-5}$) |
| (xi) | 0.30770370(8.42x$10^{-4}$) | 0.32129313(-1.06x$10^{-3}$) | 0.62899683(-1.32x$10^{-4}$) |
| (xii) | 0.30765532(-1.49x$10^{-2}$) | 0.32135340(1.77x$10^{-2}$) | 0.62900872 (1.76x$10^{-3}$) |
| (xiii) | 0.30765420(-1.52x$10^{-2}$) | 0.32135189(1.72x$10^{-2}$) | 0.62900609(1.34x$10^{-3}$) |
| (xiv) | 0.30770380(8.74x$10^{-4}$) | 0.32129328(-1.02x$10^{-3}$) | 0.62899708 (-9.22x$10^{-5}$) |
| (xv) | 0.30770113(6.50x$10^{-6}$) | 0.32129657(6.22x$10^{-6}$) | 0.62899771(7.95x$10^{-6}$) |

**Table 7**

The time period T for an AHO with a quartic potential.

| | **ω=0.2 b=1.0 C=1.0** | **ω=1.0 b=1.0 C=1.0** | **ω=1.0 b=1000 C=1000** |
|---|---|---|---|
| (i) | 7.20968477 | 4.76802201 | |
| (ii) | 7.20573281(-5.48x$10^{-2}$) | 4.76792128(-2.11x$10^{-3}$) | 0.000234364(-6.82x$10^{-2}$) |
| (iii) | 7.20551946(-5.78x$10^{-2}$) | 4.76792019(-2.14x$10^{-3}$) | 0.000234354(-7.25x$10^{-2}$) |
| (iv) | 7.21145721(2.46x$10^{-2}$) | 4.76806419(8.85x$10^{-4}$) | 0.000234596(3.07x$10^{-2}$) |
| (vi) | 7.20918046(-6.99x$10^{-3}$) | 4.76800999(-2.52x$10^{-4}$) | 0.000234503(-8.95x$10^{-3}$) |
| (vii) | 7.19776440(-0.16) | 4.76735928(-1.39x$10^{-2}$) | 0.000234065(-0.20) |
| (viii) | 7.20943604(-3.45x$10^{-3}$) | 4.76801954(-5.18x$10^{-5}$) | 0.000234513(-4.69x$10^{-3}$) |
| (ix) | 7.20852458(-1.61x$10^{-2}$) | 4.76799625(-5.40x$10^{-4}$) | 0.000234476(-2.05x$10^{-2}$) |
| (x) | 7.20844726 (-1.72x$10^{-2}$) | 4.76799540(-5.58x$10^{-4}$) | 0.000234473(-2.17x$10^{-2}$) |
| (xi) | 7.21023129(7.58x$10^{-3}$) | 4.76807964(1.21x$10^{-3}$) | 0.000234543(8.10x$10^{-3}$) |
| (xii) | 7.24825219(0.54) | 4.77319173(0.11) | 0.000235926(0.60) |
| (xiii) | 7.20974597(8.49x$10^{-4}$) | 4.76796815(-1.13x$10^{-3}$) | Failed |
| (xiv) | 7.20767032(-2.72x$10^{-2}$) | 4.76787781(-3.02x$10^{-3}$) | Failed |
| (xv) | 7.20844088(-1.73x$10^{-2}$) | 4.76798872(-6.98x$10^{-4}$) | 0.000234474(-2.13x$10^{-2}$) |
| | **ω=2.0 b=1.0 C=1.0** | **ω=5.0 b=1.0 C=1000**<br>**ω=5.0 b=100 C=100** | **ω=10.0 b=1.0 C=10000**<br>**ω=10.0 b=100 C=1000** |
| (i) | 2.88442276 | 0.00741616 | 0.000741629 |
| (ii) | 2.88442493(7.52x$10^{-5}$) | 0.00741111(-6.81x$10^{-2}$) | 0.000741124(-6.81x$10^{-2}$) |
| (iii) | 2.88442167(-3.78x$10^{-5}$) | 0.00741080(-7.23x$10^{-2}$) | 0.000741093(-7.23x$10^{-2}$) |
| (iv) | 2.88442328(1.80x$10^{-5}$) | 0.00741846(3.10x$10^{-2}$) | 0.000741858(3.09x$10^{-2}$) |
| (vi) | 2.88442264(-3.93x$10^{-6}$) | 0.00741551(-8.76x$10^{-3}$) | 0.000741564(-8.76x$10^{-3}$) |
| (vii) | 2.88440305(-6.83x$10^{-4}$) | 0.00740165(-0.20) | 0.000740177(-0.20) |
| (viii) | 2.88442276(2.53x$10^{-7}$) | 0.00741582(-4.58x$10^{-3}$) | 0.000741595(-4.58x$10^{-3}$) |
| (ix) | 2.88442250(-8.68x$10^{-6}$) | 0.00741466(-2.02x$10^{-2}$) | 0.000741478(-2.04x$10^{-2}$) |
| (x) | 2.88442250(-8.80x$10^{-6}$) | 0.00741455(-2.17x$10^{-2}$) | 0.000741468(-2.17x$10^{-2}$) |
| (xi) | 2.88442493(7.52x$10^{-5}$) | 0.00741677(8.23x$10^{-3}$) | 0.000741690(8.23x$10^{-3}$) |
| (xii) | 2.88486536(1.53x$10^{-2}$) | 0.00746049(0.60) | 0.000746063(0.60) |
| (xiii) | 2.88442033(-8.42x$10^{-5}$) | Failed | Failed |
| (xiv) | 2.88441801(-1.65x$10^{-4}$) | Failed | Failed |
| (xv) | 2.88442239(-1.28x$10^{-5}$) | 0.00741458(-2.13x$10^{-2}$) | 0.000741471(-2.13x$10^{-2}$) |

considered for the cubic AHO in Tables 5 and 6 are as per the inequality $aC < \omega^2/2$ and the cases ω = 1.0, a = 0.5, C = 0.99, and ω = 1.0, a = 1.0, C = 0.49 in Table 5 are the two extreme examples of the homoclinic orbits for the listed ω and a values as the higher values of C make the orbits heteroclinic.

It is worth noting that the expressions for $\Omega_+$ for the cubic - quartic potential AHO in all the LPM based approaches and the first–order treatments corresponding to the HBM, the EBM and the FSE contain aC and $bC^2$ as single factors and this characteristic is more prominent in the cubic as well as quartic potential AHOs where aC and $bC^2$, respectively, occur together. This aspect has been also pointed out by many earlier workers during the discussion of the quartic AHO (see e.g. [11,18]). However, this feature becomes obscure in the higher-order versions of the last three methods because C itself is taken as sum of two or more components. Furthermore, similar comments hold good for $\Omega_-$, wherein C is replaced by D. Since D itself is determined by the value of C, both the angular frequencies and thereby all the periods $T_+$, $T_-$ and T effectively depend on aC and $bC^2$. This dependence is fully collaborated by the numerical results obtained with various techniques considered in this work, for different ω values and sets of values of aC, $bC^2$ for the cubic - quartic, cubic and quartic potential AHOs. Some of such cases have been highlighted in the Tables 2, 6 and 7. For example, the results for ω = 2.0, a = 5.0, b = 100.0, C = 1.0, D = 1.03210629 and ω = 2.0, a = 0.5, b = 1.0, C = 10.0, D =10.32106290 are exactly the same and have been projected as a single entry in Table 2.

A perusal of the entries in Tables 1 - 7 shows that if the values of different AHO parameters are comparable (C is neither large nor defines the border case of the homoclinic and the heteroclinic orbits) then the PDs for the LPM based approaches are quite large for $T_+$ and $T_-$ though these become significantly less for T because their signs are opposite for the values of $T_+$ and $T_-$. Sometimes the cancellation of the errors in $T_+$ and $T_-$ is so strong that the value of T looks impressively accurate. The sets ω = 2.0, a = 1.0, b = 1.0, C = 1.0, D = 1.15167393 (Table 2); ω = 5.0, a = 1.0, b = 1.0, C = 1.0, D = 1.02630632 (Table 3); ω = 10.0, a = 0.5, b = 0.0, C = 10.0, D = 10.34523859 and the related entries (Table 6) are typical examples of such situation. Nonetheless, this does bring out the inherent limitation of the applicability of all these methods for the case of AHOs described by the asymmetric potentials. The findings of the other approaches are, generally, reasonably accurate except some versions of the EBM. Also we cannot label different methods according to their accuracy because neither this nor the technique that yields the minimum PD is unique. Of course, the PDs for the cases corresponding to the C values bordering the homoclinic and the heteroclinic orbits (e.g., C = 0.45 for ω = 2.0, a = 5.0, b = 1.0 (Table 2); C = 0.99 for ω=1.0, a = 0.5 (Table 5); and C = 0.49 in the case of ω = 1.0, a = 1.0 (Table 5)) are reasonably large for all the techniques. In particular, it may be noted that for ω = 1.0, a = 1.0, C= 0.4 (Table 5) the PDs for some methods are even of the order of $10^{-3}$, while for C = 0.49 these are higher than $10^{-1}$ and also the FSE which is the most accurate approach for the former is not so for the latter. Also some of the methods fail to yield results in these cases.

Now coming to the dependence of PDs on the value of C for specific values of other parameters, even for C = 0.1 the errors for the LPM based techniques are sufficiently large for $T_+$ and $T_-$ though these are small for T. This is not the case for other methods. For example, for $\omega = 1.0$, a = 1.0, b = 1.0, C = 0.1, the PDs in $T_+$, $T_-$ and T for the LPM Amore et al approach are, respectively, 4.55, -4.17 and $-2.4 \times 10^{-2}$ while in the case of FSE technique all the PD values are about $2 \times 10^{-7}$. As C increases, the errors generally increase. This trend has been observed for the AHOs described by the cubic - quartic, the cubic and the quartic potentials. However, when C is quite large as compared to the other parameters, the PDs attain constant values for $T_+$, $T_-$ and T corresponding to $bC^2$ becoming infinite ($>10^5$) irrespective of the individual values of $\omega$, a and b. Of course, we cannot talk about the asymptotic behaviour for a cubic potential AHO (see, for example, the entries corresponding to $\omega = 1.0$, a = 1.0, b = 1.0, C = 1000 in Table 1 and $\omega = 1.0$, a = 0.0, b = 1.0, C = 1000 in Table 7). It is found that under these conditions, the outcome of the computations for the value of T is the most reliable for the rational function based HBM followed by third-order EBM and third-order HBM. The performance of the HBM employing the rational functions proposed in this work is next in gradation and comparable with that of FSE. The LPM based techniques belong to the next group on the basis of the magnitude of PD, while the second-order HBM as well as the second-order EBM do quite poorly. Of course, the EBM using the rational function does not work. As far as appreciation of the situation pertaining to large $bC^2$ on the basis of the expressions derived in the preceding sections is concerned, this can be done for the LPM based approaches but not in other cases because these involve writing C as sum of two or more components and the moment the dominating component is taken very large ignoring the other components the technique itself is reduced to the first-order one and, therefore, its discussion becomes meaningless. In the case of Amore et al approach, for very large value of $bC^2$ ignoring the contribution of all other terms we get the same PD of $-6.82 \times 10^{-2}$ for $T_+$, $T_-$ and T of a cubic - quartic potential AHO and a quartic AHO. This result is the same as reported in [9]. Obviously, this is in complete agreement with the PD for the relevant T values in different tables, but significantly deviates from the findings for $T_+$ and $T_-$ indicating that the terms containing aC or aD do make a contribution.

It is quite interesting to investigate influence of variation in the $\omega$ values on the time periods because so far all the authors have essentially taken $\omega = 1$. When $\omega$ is much smaller than the other parameters, the PDs are reasonably large; see, e.g., the results for $\omega = 0.2$, b = 1.0, C = 1.0 in Table 7. As $\omega$ increases, the PDs become smaller particularly for the values of T and for $\omega$ nearly 10 or more times larger than the values of other parameters these differences become extremely small. In the case of a cubic - quartic potential AHO defined by a = b = C = 1.0, the data in Tables 1 - 3 for $\omega = 1.0, 2.0, 5.0$ are in accord with this statement. For $\omega = 10$, the PDs become still smaller and for $\omega = 20$, we get $T_+ = 0.15676667$, $T_- = 0.15709883$ and T = 0.31386550 in almost all the cases including the exact values implying excellent performance of the approximate methods except the LPM based techniques, the second-order EBM and the rational function based EBM where the percentage differences for $T_+$ and $T_-$ are relatively higher and are even somewhat poorer for T with the EBM approaches mentioned above. Similar comments apply to the quartic potential AHO with b = C = 1.0 and varying magnitudes of ω. In this case the results for ω = 0.2, 1.0, 2.0 have been projected in Table 7 and for ω = 5, 10 and 20, the respective values of T are 1.2382232, 0.62597623 and 0.3138651775 with exceptions as in the previous example for T values. Here, the exact values have been also rounded off to the relevant number of significant figures. In the case of a cubic potential AHO, we have used a = C = 1.0 for elaboration. In view of inequality, $aC < \omega^2/2$, ω = 1.0 has not been considered. The results for ω = 2.0 have been included in Table 6, which show that the PD is minimum for the FSE. For ω = 5, the values of $T_+$, $T_-$ and T are, respectively, 0.617921, 0.639579 and 1.257499, while for ω = 10, these are 0.3128345, 0.3155104 and 0.6283449. Once again the methods departing from the general trend are the same as in the case of a cubic - quartic potential AHO. The reason for this type of behaviour with increase in ω can be understood by realizing the fact that when the value of ω is higher than the other parameters by an order of magnitude, the latter act only as perturbation and their contribution to angular frequency is quite small which makes the results for time period accurate irrespective of the technique employed.

It is worth mentioning that if C is very large (few orders of magnitude higher than the anharmonicity parameters) for cubic - quartic and quartic potential AHOs, the PDs are not affected by increasing the ω values by a factor of 10 or 20. For example, for ω = 10.0, a = 1.0, b = 1.0, C = 1000, D = 1000.6666002 these are the same as for ω = 1.0, a = 1.0, b = 1.0, C = 1000, D = 1000.66666615 given in Table 1. This behaviour is a consequence of the dominance of the large value of $bC^2$.

In the case of a quartic potential AHO, some workers [11,18] have evaluated the angular frequency or time period for the negative values of $bC^2$, which essentially implies b < 0. Here, we have examined in some detail the effect of change in sign of the parameters a as well as b.

If we replace a by – $a_1$ in the potential for a cubic AHO, the expression becomes

$$V(q) = m\left[\frac{1}{2}\omega^2 q^2 - \frac{1}{3}a_1 q^3\right]. \tag{98}$$

Obviously, the curve for q(t) > 0 will match the one for q(t) < 0 in the cubic AHO with positive a so that the plot for V(q) corresponding to negative a is mirror image in the Y-axis of that for the potential with positive a. Accordingly, the values of C and D and hence those of $T_+$ and $T_-$ get interchanged and the total time period T is unaffected.

For a quartic potential AHO, changing b to – $b_1$ gives

$$V(q) = m\left[\frac{1}{2}\omega^2 q^2 - \frac{1}{4}b_1 q^4\right], \tag{99}$$

which has $q_\pm = \frac{\pm\omega}{\sqrt{b_1}}$ as two maxima with $q_0 = 0$ as a stable minimum in between. In this case, the amplitude of oscillations has to be such that

$$0 < C < \omega/\sqrt{|b|}. \tag{100}$$

Some representative results for the quartic AHO with negative b values constitute the content of Table 8. A look at these results shows that the PDs increase as $|b|C^2$ approaches $\omega^2$. Here too, if ω is increased for a particular set of b and C values, the PDs decrease. For example, for b = - 1.0 and C = 1.0 when ω is increased from 2.0 (for which the results are listed in Table 8) to 10.0 the time periods obtained with all the techniques including the exact result can be rounded off to 0.63068881 except that determined by EBM (second-order) which yields the value as 0.63068904. Furthermore, a comparison of the findings for ω = 2.0, b = - 1.0, C = 1.0 (Table 8) with those for ω = 2.0, b = 1.0, C = 1.0 (Table 7) reveals that the time periods for the negative b are higher and the PDs also get enhanced even though the relative accuracy of various techniques is essentially the same.

**Table 8**

Time periods T for an AHO governed by quartic potential, with negative value of the parameter b and the legends for numbers (i) to (xv) the same as in Table 1.

| | **ω=1.0 b= -1.0 C=0.4** | **ω=1.0 b= -1.0 C=0.95** |
|---|---|---|
| (i) | 6.70049566 | 12.47973673 |
| (ii) | 6.70049430($-2.03\times10^{-5}$) | 12.22757392(-2.02) |
| (iii) | 6.70049429($-2.04\times10^{-5}$) | 12.16677725(-2.51) |
| (iv) | 6.70049631($9.74\times10^{-6}$) | 12.68611211(1.65) |
| (vi) | 6.70049555($-1.61\times10^{-6}$) | 12.43071039(-0.39) |
| (vii) | 6.70052538($4.44\times10^{-4}$) | 13.12769105(5.19) |
| (viii) | 6.70049573($1.10\times10^{-6}$) | 12.52368388(0.35) |
| (ix) | 6.70049539($-3.95\times10^{-6}$) | 12.41723016(-0.50) |
| (x) | 6.70049540($-3.89\times10^{-6}$) | 12.43420184(-0.36) |
| (xi) | 6.70049169($-5.92\times10^{-5}$) | 12.30638683(-1.39) |
| (xii) | 6.70132588($1.24\times10^{-2}$) | Failed |
| (xiii) | 6.70050077($7.63\times10^{-5}$) | 12.93049048(3.61) |
| (xiv) | 6.70050362($1.19\times10^{-4}$) | 12.69575555(1.73) |
| (xv) | 6.70049521($-6.75\times10^{-6}$) | 12.29080229(-1.51) |
| | **ω=2.0 b= -1.0 C=1.0** | **ω=5.0 b= -100 C=0.40** |
| (i) | 3.48916347 | 1.77529054 |
| (ii) | 3.48915738($-1.75\times10^{-4}$) | 1.77442539($-4.87\times10^{-2}$) |
| (iii) | 3.48915736($-1.75\times10^{-4}$) | 1.77438126($-5.12\times10^{-2}$) |
| (iv) | 3.48916603($7.32\times10^{-5}$) | 1.77567662($2.17\times10^{-2}$) |
| (vi) | 3.48916277($-2.02\times10^{-5}$) | 1.77518083($-6.18\times10^{-3}$) |
| (vii) | 3.48923889($2.16\times10^{-3}$) | 1.77812661(0.16) |
| (viii) | 3.48916357($2.72\times10^{-6}$) | 1.77533314($2.40\times10^{-3}$) |
| (ix) | 3.48916211($-3.93\times10^{-5}$) | 1.77510547($-1.04\times10^{-2}$) |
| (x) | 3.48916213($-3.84\times10^{-5}$) | 1.77512223($-9.48\times10^{-3}$) |
| (xi) | 3.48915258($-3.12\times10^{-4}$) | 1.77468023($-3.44\times10^{-2}$) |
| (xii) | 3.49044360($3.67\times10^{-2}$) | 1.79081835(0.87) |
| (xiii) | 3.48917746($4.01\times10^{-4}$) | 1.77615402($4.86\times10^{-2}$) |
| (xiv) | 3.48918416($5.93\times10^{-4}$) | 1.77621879($5.23\times10^{-2}$) |
| (xv) | 3.48916124($-6.41\times10^{-5}$) | 1.77490790($-2.16\times10^{-2}$) |

A replacement of both a and b by – $a_1$ and – $b_1$, respectively, in the expression for potential of a cubic-quartic AHO transforms this into

$$V(q) = m\left[\frac{1}{2}\omega^2 q^2 - \frac{1}{3}a_1 q^3 - \frac{1}{4}b_1 q^4\right]. \quad (101)$$

This has $q_0$ as a stable minimum and $q_{\pm} = \frac{-a_1 \pm (a_1^2+4\omega^2 b_1)^{\frac{1}{2}}}{2b_1}$ as local maxima. The plot for V(q) in this case is also a mirror image in the Y-axis of that for the potential with positive values of a and b. However, the initial amplitude C of the oscillations (q(t) > 0) should satisfy the condition $C < \frac{-a_1 \pm (a_1^2+4\omega^2 b_1)^{\frac{1}{2}}}{2b_1}$ . The equation defining the amplitude D for the motion corresponding to q(t) < 0 for a specific value of C reads

$$\frac{1}{4} b_1 D^4 - \frac{1}{3} a_1 D^3 - \frac{1}{2}\omega^2 D^2 = \frac{1}{4} b_1 C^4 + \frac{1}{3} a_1 C^3 - \frac{1}{2}\omega^2 C^2. \tag{102}$$

Once again, the values of C and D and hence those of $T_+$ and $T_-$ for the potential having negative a and b are interchanged with respect to the ones for the potential with positive a and b and, therefore, the total time period T is the same.

The purely cubic-quartic potential AHO (ω = 0.0) too has been included in the present comparative study. It reads

$$V(q) = m\left[\frac{1}{3} a q^3 + \frac{1}{4} b q^4\right] \tag{103}$$

and has $q_0$ = - a/b as a local minimum and q = 0 as an inflection. Shifting the point of reference to $q_0$ = - a/b by substituting q'= q + a/b, Eq. (103) is transformed into

$$V(q') = m\left[\frac{1}{2}\omega'^2 q'^2 + \frac{1}{3} a' q'^3 + \frac{1}{4} b' q'^4 - \frac{a'^4}{192\, b'^3}\right], \tag{104}$$

where $\omega'^2 =$ $a^2$/b, a' = - 2a and b' = b. In this case the amplitude C = a/b corresponds to infinite time period, while the magnitudes of C less or greater than this lead to finite periods. It may be noted that ω' is large for the high values of a and small values of b. Also since the slope for q(t) > 0 is less than that for q(t) < 0, D < C. The results of computations for some typical values of the parameters for the purely cubic - quartic AHO are listed in Table 9.

**Table 9**

The time periods for q(t) > 0, q(t) < 0 and total for a purely cubic - quartic potential AHO (ω = 0.0).

| | $T_+$ | $T_-$ | T |
|---|---|---|---|
| **a=0.5 b=0.5 C=0.5 D=0.28414187** | | | |
| (i) | 7.76185565 | 3.58041047 | 11.34226611 |
| (ii) | 5.18639466(-33.18) | 5.14532900(43.71) | 10.33172366(-8.91) |
| (iii) | 4.21997634(-45.63) | 5.14283018(43.64) | 9.36280652(-17.41) |
| (v) | - | - | 11.21338561(-1.14) |
| (vi) | 7.75842014(-4.43x10$^{-2}$) | 3.58038055(-8.36x10$^{-4}$) | 11.33880069(-3.06x10$^{-2}$) |
| (vii) | 7.78473270(0.29) | 3.58059566(5.17x10$^{-3}$) | 11.36532836(0.20) |
| (viii) | 7.77580312(0.18) | 3.58066282(7.05x10$^{-3}$) | 11.35646594(0.13) |
| (ix) | 7.75777355(-5.26x10$^{-2}$) | 3.58038149(-8.09x10$^{-4}$) | 11.33815505(-3.62x10$^{-2}$) |
| (x) | 7.759943392(-3.12x10$^{-2}$) | 3.58037524(-9.84x10$^{-4}$) | 11.33980916(-2.17x10$^{-2}$) |
| (xi) | 7.74471282(-0.22) | 3.58054385(3.73x10$^{-3}$) | 11.32525667(-0.15) |
| (xii) | 7.90701645(1.87) | 3.57731332(-8.65x10$^{-2}$) | 11.48432977(1.25) |
| (xiii) | 7.84504696(1.07) | 3.57621856(-0.12) | 11.42126552(0.70) |
| (xiv) | 7.75797598(-5.00x10$^{-2}$) | 3.58060450(5.42x10$^{-3}$) | 11.33858048(-3.24x10$^{-2}$) |
| (xv) | 7.76011987(-2.24x10$^{-2}$) | 3.58042696(4.61x10$^{-4}$) | 11.34054683(-1.52x10$^{-2}$) |
| **a=0.5 b=0.5 C=1000 D=998.66666682** | | | |
| (i) | 0.00525028 | 0.00524494 | 0.01049523 |
| (ii) | 0.00524339(-0.13) | 0.00524471(-4.39x10$^{-3}$) | 0.01048810(-6.79x10$^{-2}$) |
| (iii) | 0.00524293(-0.14) | 0.00524465(-5.53x10$^{-3}$) | 0.01048758(-7.29x10$^{-2}$) |
| (iv) | 0.00525111(1.58x10$^{-2}$) | 0.00524742(4.73x10$^{-2}$) | 0.01049853(3.14x10$^{-2}$) |
| (v) | - | - | 0.01049831(2.93x10$^{-2}$) |
| (vi) | 0.00524982(-8.76x10$^{-3}$) | 0.00524448(-8.77x10$^{-3}$) | 0.01049431(-8.77x10$^{-3}$) |
| (vii) | 0.00523996(-0.20) | 0.00523472(-0.19) | 0.01047468(-0.20) |
| (viii) | 0.00525003(-4.76x10$^{-3}$) | 0.00524472(-4.19x10$^{-3}$) | 0.01049475(-4.57x10$^{-3}$) |
| (ix) | 0.00524921(-2.04x10$^{-2}$) | 0.00524388(-2.02x10$^{-2}$) | 0.01049309(-2.04x10$^{-3}$) |
| (x) | 0.00524914(-2.17x10$^{-2}$) | 0.00524381(-2.15x10$^{-2}$) | 0.01049295(-2.17x10$^{-2}$) |
| (xi) | 0.00525072(8.38x10$^{-3}$) | 0.00524537(8.20x10$^{-3}$) | 0.01049609(8.20x10$^{-3}$) |
| (xii) | 0.00528178(0.60) | 0.00527619(0.60) | 0.01055797(0.60) |
| (xv) | 0.00524916(-2.13x10$^{-2}$) | 0.00524383(-2.12x10$^{-2}$) | 0.01049298(-2.14x10$^{-2}$) |
| **a=100.0 b=1.0 C=20.0 D=15.70492113** | | | |
| (i) | 0.03785280 | 0.02772317 | 0.06557597 |
| (ii) | 0.03267196(-13.69) | 0.03266315(17.82) | 0.06533511(-0.37) |
| (iii) | 0.03234569(-14.55) | 0.03264858(17.77) | 0.06499427(-0.89) |
| (iv) | 0.03374101(-10.86) | 0.03249539(17.21) | 0.06623640(1.01) |
| (v) | - | - | 0.06550478(-0.11) |
| (vi) | 0.03785213(-1.77x10$^{-3}$) | 0.02772307(-3.57x10$^{-4}$) | 0.06557521(-1.16x10$^{-3}$) |
| (vii) | 0.03785769(1.29x10$^{-2}$) | 0.02772382(2.34x10$^{-3}$) | 0.06558151(8.45x10$^{-3}$) |
| (ix) | 0.03785210(-1.85x10 $^{-3}$) | 0.02772307(-3.61x10$^{-4}$) | 0.06557517(-1.22x10$^{-3}$) |
| (x) | 0.03785221(-1.56x10$^{-3}$) | 0.02772306(-3.97x10$^{-4}$) | 0.06557528(-1.05x10$^{-3}$) |
| (xi) | 0.03784618(1.75x10$^{-2}$) | 0.02772406(3.21x10$^{-3}$) | 0.06557023(-8.75x10$^{-3}$) |
| (xii) | 0.03793916(0.23) | 0.02770487(-6.60x10$^{-2}$) | 0.06564403(0.10) |
| (xv) | 0.03785297(4.49x10$^{-4}$) | 0.02772322(1.80x10$^{-4}$) | 0.06557619(3.35x10$^{-4}$) |

It is observed that for a specific set of a, b and C (not very large) values for $\omega = 0$, the periods determined with the approximate techniques differ significantly from the corresponding exact value as compared to the case when $\omega$ is nonzero; see, e.g., entries for a = b = C = 0.5 in Tables 1 and 9. However, if C is taken quite large keeping a as well as b small, the PDs generally approach the values found with nonzero $\omega$. The LPM based techniques are exceptions for quite small a and b but not for the large values. As pointed out above $\omega'$ becomes large when a is large and this, in turn, leads to accurate results for $\omega'$ much larger than C, as found in the earlier cases. Such findings have been included in Table 9 for a = 100.0, b = 1.0, C = 20.0, while for C = 10 and 5, the PDs turn out to be sufficiently small.

It was contended after Eq. (67) that the use of the rational function in HBM should yield results better than the second-order HBM but worse than the third-order HBM. A consequence of the argument given there is that the two higher-order rational functions proposed in the present work should improve the situation even over the third-order HBM. In a similar manner, the outcome of the EBM based on these rational functions should also be in accord with the above statement. However, in practice, this is not always true for different values of the AHO parameters as can be seen from the content of different tables. For large value of $bC^2$, the magnitudes of PDs for HBM are in agreement with the assertion made here though the value for the second higher-order rational function is an exception.

**8. Concluding remarks**

Guided by the wide range applications of the AHOs characterized by the non-perturbative parameters and the methods put forward to determine approximate values of their angular frequency and hence time period, we have performed a detailed comparative study of the different modified versions of the Lindstedt-Poincare technique, the harmonic balance method based on the usual approach and on the use of rational functions , the energy balance method and the two–terms Fourier series technique by considering the cubic - quartic potential AHO (and its special cases of cubic, quartic and purely cubic - quartic AHOs) as the test case. The conditions for the orbits to be homoclinic in the asymmetric potential AHOs have been obtained and used to define the initial value of the amplitude of oscillations C for q(t )> 0. These, in turn, have been employed to determine the corresponding amplitude D for q(t) < 0. Two new rational functions as approximate solutions in HBM have been put forward. Also the rational functions as trial functions in the EBM have been used for the first time.

The analysis has been carried out by deriving relationships between angular frequency for q(t) > 0 and q(t) < 0 and various AHO parameters ω, a, b and C or D corresponding to different techniques. These have been then employed to determine the relevant time periods $T_+$, $T_-$ and the total time period T for a wide spectrum of the parameter values, including the negative ones of b. The results have been compared with the relevant exact values and have been thoroughly assessed with respect to their different features, viz., the occurrence of aC and $bC^2$ together, the influence of variation in C from small values to very large values including those bordering the homoclinic and heteroclinic orbits, the effect of increase in the value of ω, and change in the sign of the anharmonicity parameters a and b. The method for which the PD of an approximate period from the corresponding exact value is the least is not the same in all the cases considered here. Nonetheless, in general, the LPM based approaches have been found to be inadequate in the sense that these do not yield accurate results for $T_+$ and $T_-$. The overall performance of the HBM based techniques is the most superior particularly keeping in view the relatively less effort involved in their implementation. The two-term FSE is the next which is followed by the EBM. It may be pointed out that the conclusion in respect of the third-order EBM is in contrast with the claim by Durmaz and Kaya [20] that its performance is better than that of the HBM. Moreover, EBM using rational functions has not come up to the mark as this failed in some cases.